\documentclass[twocolappendix]{emulateapj}  

\usepackage{color}
\usepackage{epsfig}
\usepackage{subfigure,verbatim}
\usepackage[colorlinks,urlcolor=blue,citecolor=blue,linkcolor=blue]{hyperref}
 \usepackage{hhline} 

\newcommand{\unit}[1]{\nobreak{\mathrm{\;#1}}} 
\newcommand{\ud}{{\rm d}} 

\newcommand{\grad}{\nabla}
\newcommand{\cross}{\times}
\newcommand{\bmath}[1]{\mbox{\boldmath{$#1$}}}

\newcommand{\bi}{\begin{itemize}}
\newcommand{\ei}{\end{itemize}}

\newcommand{\ex}[1]{10^{-#1}}

\newcommand{\eq}[1]{Eq.~(\ref{eq:#1})}
\newcommand{\eqn}[1]{(\ref{eq:#1})}
\newcommand{\fig}[1]{Fig.~\ref{fig:#1}}
\newcommand{\fign}[1]{\ref{fig:#1}}
\newcommand{\tab}[1]{Table~\ref{tab:#1}}
\newcommand{\sect}[1]{Section \ref{sec:#1}}
\newcommand{\app}[1]{Appendix \ref{sec:#1}}
\newcommand{\dg}{\Delta\gamma}
\newcommand{\be}{\begin{eqnarray}}
\newcommand{\ee}{\end{eqnarray}}

\newcommand{\qt}{(1+q\,t)}
\newcommand{\bavg}{\langle \bmath{B}\rangle}

\newcommand{\perpe}{{e\perp}}
\newcommand{\pare}{{e\parallel}}
\newcommand{\perpi}{{i\perp}}
\newcommand{\pari}{{i\parallel}}

\def\L{\bmath{{L}}}
\def\bvec{\bmath{B}}

\def\evec{\bmath{E}}

\def\ra{\rangle}
\def\la{\langle}
\def\powi{{0.4}}
\def\powe{{0.5}}
\def\coeffi{{0.65}}
\def\coeffe{{0.55}}

\begin{document}
\title{Electron Heating by the Ion Cyclotron Instability  in Collisionless Accretion Flows.\\
II. Electron Heating Efficiency as a Function of Flow Conditions}
\author{Lorenzo Sironi$^{1,2}$}
\affil{$^1$Harvard-Smithsonian Center for Astrophysics, 
60 Garden Street, Cambridge, MA 02138, USA
\\
$^2$NASA Einstein Postdoctoral Fellow}
 \email{E-mail: lsironi@cfa.harvard.edu}
 
\begin{abstract}
In the innermost regions of low-luminosity accretion flows, including Sgr A$^*$ at the center of our Galaxy, the frequency of Coulomb collisions is so low that the plasma is two-temperature, with the ions substantially hotter than the electrons. This paradigm assumes that Coulomb collisions are the only channel for transferring the ion energy to the electrons. In this work, the second of a series, we assess the efficiency of electron heating by ion velocity-space instabilities  in collisionless accretion flows. The instabilities are seeded by the pressure anisotropy induced by magnetic field amplification, coupled to the adiabatic invariance of the particle magnetic moments. Using two-dimensional (2D) particle-in-cell (PIC) simulations, we showed in Paper I that  if the electron-to-ion temperature ratio is $T_{0e}/T_{0i}\lesssim0.2$, the ion cyclotron instability is the dominant mode for ion betas $\beta_{0i}\sim 5-30$ (here, $\beta_{0i}$ is the ratio of ion thermal pressure to magnetic pressure), as appropriate for the midplane of low-luminosity accretion flows. In this work, we employ analytical theory and 1D PIC simulations (with the box aligned with the fastest growing wavevector of the ion cyclotron mode) to fully characterize how the electron heating efficiency during the growth of the ion cyclotron instability depends on the electron-to-proton temperature ratio, the plasma beta, the Alfv\'en speed, the amplification rate of the mean field (in units of the ion Larmor frequency) and the proton-to-electron mass ratio. Our findings can be incorporated as a physically-grounded sub-grid model into global fluid simulations of low-luminosity accretion flows, thus helping to assess the validity of the two-temperature assumption. 
\end{abstract}

 \shorttitle{Electron Heating by the Ion Cyclotron Instability  in Collisionless Accretion Flows}
\shortauthors{L. Sironi}


\section{Introduction}
In low-luminosity accretion flows,  including the
ultra-low-luminosity source Sagittarius A$^*$ (Sgr A$^*$) at our
Galactic Center
\citep{nym95,nar+98,yuan+03,xu+06,moscibrodzka+12,yuan_narayan_14}, the timescale for electron and ion Coulomb collisions is much longer than the inflow time in the disk, i.e., the plasma is collisionless. At distances less than a few hundred
Schwarzschild radii from the black hole ($R_S\equiv 2\,GM_\bullet/c^2$ is the Schwarzschild radius, where $M_\bullet$ is the black hole mass), ions and electrons are thermally decoupled and the plasma is two-temperature, with
the ions substantially hotter than the electrons
\citep{ny95b,yuan+03}. This stems from the fact that (\textit{i}) compressive heating favors non-relativistic ions over relativistic electrons, and that (\textit{ii}) electrons copiously lose energy via radiative cooling.

Early work on two-temperature models of low-luminosity accretion flows (also known as ADAFs, or advection-dominated accretion flows) assumed that
most of the turbulent viscous energy goes into the ions
\citep{ichimaru77,rees82,ny95b}, and that only a small fraction $\sim
10^{-3}-10^{-2}$ goes into the electrons.  There have been occasional
attempts to estimate this fraction from microphysics, by considering
magnetic reconnection \citep{bisno97,quataertgruzinov99}, magnetohydrodynamic (MHD)
turbulence \citep{quataert98,blackman99,medvedev00}, plasma waves  \citep{begelman88}, or dissipation of
pressure anisotropy in collisionless plasmas
\citep{sharma07}. Currently, it appears that the electron-to-proton temperature ratio lies
 in the range $0.1-0.5$, but this is by and large no more
than a guess \citep{ny95b,yuan+03,yuan_narayan_14}.\footnote{Note  that the plasma in the fast solar wind and behind shocks in
  supernova remnants is also two-temperature
  \citep{marsch12,rakowski05,ghavamian07,morlino_12}.} The micro-physics of energy dissipation and electron heating  in collisionless accretion flows cannot be captured in the MHD framework, but it requires a fully-kinetic description with first-principles particle-in-cell (PIC) simulations.
  
In this work, the second of a series, we study with PIC simulations the efficiency of electron heating by ion velocity-space instabilities, in the context of low-luminosity accretion disks. Pressure anisotropies are continuously generated in collisionless accretion flows due to the fluctuating magnetic fields associated with the non-linear stages of the magnetorotational instability (MRI, \citealt{balbus91,balbus98}),
a MHD instability that governs the transport of angular momentum in accretion disks \citep[see][for a study of the collisionless MRI]{riquelme12,hoshino_13}. If the magnetic field is amplified by the MRI, the adiabatic invariance of the magnetic moments of charged particles drives the   
perpendicular (to the
magnetic field) pressure $P_\perp$ to be much greater than the parallel
pressure $P_\parallel$, a configuration prone to velocity-space instabilities. The same instabilities are believed to play an important role in the solar wind \citep{kasper_02,kasper_06,bale_09,maruca_11,maruca_12,matteini_07,matteini_13,cranmer_09,cranmer_12} and in the intracluster medium \citep{scheko_05,lyutikov_07,santos-lima_14}.

  In Paper I, we have developed a fully-kinetic method for studying velocity-space instabilities in a system where the field is continuously amplified. In this case, the anisotropy is constantly driven (as a result of the field amplification), rather than assumed as a prescribed initial condition, as in most earlier works \citep[see][for a review]{gary_book}. In our setup, the increase in magnetic field is driven by {\it compression} \citep[as in][]{hellinger_05}\footnote{Note that \citet{hellinger_05}  employed a hybrid code --- that treats the ions as kinetic particles, but the electrons as a massless charge-neutralizing fluid. Thus, the electron kinetic physics was not properly captured.},  mimicking the effect of large-scale compressive motions in ADAFs.  However, our results hold regardless of what drives the field amplification, so they can be equally applied to the case where velocity-space instabilities are induced by incompressible {\it shear} motions \citep[as in][]{riquelme_14,kunz_14}.
  
Most of the previous numerical
studies of anisotropy-driven
instabilities  focused either on ion instabilities alone (with hybrid
codes, see \citealt{gary_book,hellinger_06} for a review; or fully-kinetic simulations in electron-positron plasmas,  \citealt{riquelme_14}) or on electron modes alone
(keeping the ions as a static neutralizing background, see \citealt{gary_book} for a review). Neither
approach can capture self-consistently the energy transfer from ions to electrons, which
requires fully-kinetic simulations with mobile ions and a realistic
mass ratio. This is the purpose of our work.

With fully-kinetic PIC simulations, we study the efficiency of electron heating that results from ion velocity-space instabilities driven by magnetic field amplification in collisionless accretion flows. We focus on the regime $T_{0e}/T_{0i}\lesssim 0.2$ (here, $T_{0e}$ and $T_{0i}$ are the electron and ion temperatures, respectively) where, as demonstrated in Paper I, the dominant mode is the ion cyclotron instability \citep[e.g.,][]{gary_76,gary_93,hellinger_06}, rather than the mirror instability \citep[e.g.,][]{hasegawa_69,southwood_93,kivelson_96}. Since the wavevector of the ion cyclotron instability is along the mean magnetic field, the relevant physics can be conveniently studied by means of 1D simulations, with the box aligned with the ordered field. Via 1D PIC
simulations, we assess here the dependence of the electron heating
efficiency on the initial ratio
between electron and proton temperatures, the ion beta $\beta_{0i}$ (namely, the ratio of ion thermal pressure to magnetic pressure), the Alfv\'en speed, the amplification rate of the mean field (in units of the ion Larmor frequency) and the proton-to-electron mass ratio. Eqs.~\eqn{scal1}-\eqn{scal3} emphasize how the various  contributions to electron heating depend on the flow conditions. Their sum gives the overall electron energy gain due to the growth of the ion cyclotron instability, and it represents the main result of this paper.

This work is organized as follows. In \sect{cond}, we describe the physical conditions in the innermost regions of low-luminosity accretion flows, where the plasma is  believed to be two-temperature. The setup of our simulations is discussed in \sect{setup}. \sect{heating} summarizes the conclusions of Paper I and anticipates the 
main results of this work (consisting of Eqs.~\eqn{scal1}-\eqn{scal3}), which are extensively substantiated by our findings in \sect{flow}. We discuss the astrophysical implications of our work in \sect{summary}.

\section{Physical Conditions in Accretion Disks}\label{sec:cond}
For our investigation of velocity-space instabilities in collisionless accretion flows, we employ values of the ion temperature $T_{0i}$, the ion beta $\beta_{0i}=8 \pi n_0 k_B T_{0i}/B_0^2$ and the Alfv\'en velocity $v_{A0i}=B_0/\sqrt{4 \pi n_0 m_i}$  that are consistent with estimates taken from state-of-the-art general-relativistic magnetohydrodynamic (GRMHD) simulations of Sgr A$^*$, the low-luminosity accretion flow at our Galactic Center \citep[e.g.,][]{sadowski_13}. Here, $n_0$ and $B_0$ are the ion number density and the magnetic field strength, respectively.

GRMHD simulations can successfully describe the linear and non-linear evolution of the MRI. Yet, they cannot properly capture the dissipation of MRI-driven turbulence on small scales (the electron Larmor radius), which ultimately regulates the temperature balance between protons and electrons, in the innermost regions of ADAFs  where the plasma is two-temperature. It follows that GRMHD simulations cannot predict the value of the electron-to-proton temperature ratio $T_{0e}/T_{0i}$ to be employed in our PIC experiments, so we will have to study the dependence of our results on this parameter.

We estimate the values of $T_{0i}$, $\beta_{0i}$ and $v_{A0i}$ in the innermost regions of ADAFs  (at distances less than a few hundred Schwarzschild radii from the black hole) from Fig.~1 of \citet{sadowski_13}. The ion temperature decreases with distance from the black hole as $R^{-1}$, from $T_{0i}\sim 10^{11.5}\unit{K}$ at the innermost stable circular orbit ($R_{\rm ISCO}=3\, R_{S}$, for a non-rotating black hole) down to  $T_{0i}\sim 10^{10}\unit{K}$ at $R\sim 100 \,R_S$. In units of the proton rest mass energy, it varies in the range $k_B T_{0i}/m_i c^2\sim 0.001-0.03$, i.e., the ions are non-relativistic. On the other hand, if the electrons were to be in equipartition with the ions, they would have $k_B T_{0e}/m_e c^2\sim 1.8-50$, i.e., the electrons can be relativistically hot. 

The ion beta is nearly constant with radius, and along the disk midplane it ranges from $\beta_{0i}\sim 10$ to $\beta_{0i}\sim 30$ (Fig.~1 of \citet{sadowski_13}).\footnote{GRMHD simulations are only sensitive to the total plasma beta, including ions and electrons. For the estimates presented in this section, we implicitly assume that electrons do not contribute much to the total plasma pressure.} Lower values of the plasma beta are expected at high latitudes above and below the disk, in the so-called corona, where the plasma might be magnetically dominated (i.e., $\beta_{0i}\ll1$). In this work, we only focus on velocity-space instabilities triggered in the bulk of the disk, and we take $\beta_{0i}=5$ as the lowest value we consider, arguing that our results can be applied all the way down to $\beta_{0i}\sim 1$. We extend our investigation to higher values of $\beta_{0i}$, up to $80$. Most likely, this is not directly relevant for low-luminosity accretion flows, but it might have important implications for the physics of the intracluster medium \citep{scheko_05,lyutikov_07,santos-lima_14}.

The value of the Alfv\'en velocity can be derived from the ion temperature and the ion beta, since $v_{A0i}/c=\sqrt{2 k_B T_{0i}/\beta_{0i} m_i c^2}$. It follows that the Alfv\'en speed varies in the range $v_{A0i}/c\sim 0.01-0.1$, if $k_B T_{0i}/m_i c^2\sim 0.001-0.03$ and $\beta_{0i}\sim 10-30$. Below, we show that our results are not sensitive to variations in the Alfv\'en velocity, in the range $v_{A0i}/c\sim 0.025-0.1$ that we explore.

To fully characterize our system, we need to specify the rate $q$ of magnetic field amplification, in units of the ion gyration frequency $\omega_{0ci}=eB_0/m_i c$.  In accretion flows, we expect the ratio $\omega_{0 ci}/q$ to be much larger than unity ($\sim 10^7$, if $q$ is comparable to the local orbital frequency). For computational convenience, we employ smaller values of $\omega_{0 ci}/q$, exploring the range $\omega_{0 ci}/q=50-3200$, and we show that our results converge in the limit $\omega_{0 ci}/q\gg1$. 

Also, due to computational constraints, fully-kinetic PIC simulations are forced to employ a reduced value of the ion-to-electron mass ratio (e.g., \citealt{riquelme_14} primarily studied velocity-space instabilities in electron-positron plasmas, i.e., with $m_e=m_i$). Most of our results employ a reduced mass ratio ($m_i/m_e=16$ or 64). However, we show that our conclusions can be readily rescaled up to realistic mass ratios (we extend our study up to $m_i/m_e=1024$). This is extremely important for low-luminosity accretion disks, since only with a realistic mass ratio one can describe the case of non-relativistic ions and ultra-relativistic electrons that is of particular interest for ADAF models.

\section{Simulation Setup}\label{sec:setup}
We investigate electron heating due to ion velocity-space instabilities in collisionless accretion flows by means of fully-kinetic PIC simulations. We have modified the three-dimensional (3D) electromagnetic PIC code TRISTAN-MP \citep{buneman_93,spitkovsky_05,ssa_13,ss_14,sironi_giannios_14} to account for the effect of an overall {\it compression} of the system.\footnote{Our method can also be applied to an expanding plasma, but in this work we only study compressing systems.} Our model is complementary to the technique described in  \citet{riquelme12} and employed in \citet{riquelme_14}, which is appropriate for incompressible {\it shear} flows.

Our method, the first of its kind for fully-kinetic PIC simulations of compressing systems, has been extensively described in Paper I (there, see Section 2 and Appendix A). For completeness, we report here its main properties. We  solve Maxwell's equations and the Lorentz force in the fluid {\it comoving} frame, which is related to the {\it laboratory} frame by a Lorentz boost. In the comoving frame, we define two sets of spatial coordinates, with the same time coordinate. The {\it unprimed} coordinate system has a basis of unit vectors, so it is the appropriate coordinate set to measure all physical quantities. Yet,  we find it convenient to re-define the unit length of the spatial axes in the comoving frame such that a particle subject only to compression stays at fixed coordinates. This will be our {\it primed} coordinate system. 

The location of a particle in the laboratory frame (identified by the subscript ``L'') is related to its position in the primed coordinate system of the fluid comoving frame by $\bmath{x}_{\rm L}= \bmath{L}\, \bmath{x}'$, where compression is accounted for by the diagonal matrix 
\be
\L=\frac{\partial \bmath{x}}{\partial \bmath{x}'}=
\left(
\begin{array}{ccc}
1 \,\,\,& 0 \,& 0 \\
 0\,\,\,& \qt^{-1} \,& 0 \\
 0 \,\,\,& 0 \,& \qt^{-1}\\
\end{array}\right)~~~,
\ee
which describes compression along the $y$ and $z$ axes. By defining the determinant $\ell={\rm det}(\L)=\qt^{-2}$, the two evolutionary Maxwell's equations of the PIC method in a compressing box are, in the limit $|\dot{\L}\, \bmath{x}'|/c\ll1$ of non-relativistic compression speeds,
\be
\nabla'\cross(\L \bmath{E})  &=&-\frac{1}{c}\frac{\partial}{\partial t'}(\ell\, \bmath{L}^{-1} \bmath{B})~~,\label{eq:basic1}\\
\nabla'\cross(\L \bmath{B}) &=&\frac{1}{c}\frac{\partial}{\partial t'}(\ell\, \bmath{L}^{-1} \bmath{E})+\frac{4\pi}{c} \ell\, \bmath{J}'~~,\label{eq:basic2}
\ee
where the temporal and spatial derivatives pertain to the primed coordinate system (the reader is reminded that the primed and unprimed systems share the same time coordinate, so $\partial/\partial t'=\partial/\partial t$, whereas the spatial derivatives differ: $\nabla'=\L \,\nabla$). We define  $\evec$ and $\bvec$ to be the physical electromagnetic fields measured in the unprimed coordinate system. The current density $\bmath{J}'$ is computed by summing the contributions of individual particles, as we specify in Appendix A of Paper I.  

The equations describing the motion of a particle with charge $q$ and mass $m$ can be written, still in the limit $|\dot{\L}\, \bmath{x}'|/c\ll1$ of non-relativistic compression speeds, as
\be
\frac{\ud \bmath{p}}{\ud t'}&=&-\,\dot{\L} \L^{-1} \bmath{p}+q\left(\evec+\frac{\bmath{v}}{c}\cross \bvec\right)\label{eq:basic3}~~,\\
\frac{\ud \bmath{x'}}{\ud t'}&=&\bmath{v}'~~,\label{eq:basic4}
\ee
where $\dot{\L}=\ud \L/\ud t$. The  physical momentum $\bmath{p}=\gamma m \bmath{v}$ and velocity $\bmath{v}$ of the particle are measured in the unprimed coordinate system (here, $\gamma=1/\sqrt{1-(\bmath{v}/c)^2}$ is the particle Lorentz factor). Yet, the particle velocity $\bmath{v}'$ entering \eq{basic4} refers to the primed coordinate system, where $\bmath{v}'=\L^{-1} \bmath{v}$. Eqs.~\eqn{basic3} and \eqn{basic4} hold for particles of arbitrary Lorentz factor.

A uniform ordered magnetic field $\bmath{B}_0$ is initialized along the $x$ direction. As a result of compression, \eq{basic1} dictates that it should grow in time as $\bvec=\bvec_0\qt^2$, which is consistent with flux freezing (the particle density in the box increases at the same rate). From the Lorentz force in \eq{basic3}, the component of particle momentum aligned with the field does not change during compression, so $p_{\parallel}=p_{0\parallel}$, whereas the perpendicular momentum increases as $p_\perp=p_{0\perp}\qt$. This is consistent with the conservation of the first ($\mu\propto p_{\perp}^2/| \bvec|$) and second ($J\propto p_{\parallel} |\bvec|/n$) adiabatic invariants.

Our computational method is implemented for 1D, 2D and 3D computational domains. We use periodic boundary conditions in all directions, assuming that the system is locally homogeneous, i.e., that gradients in the density or in the ordered field $\bvec_0$ are on scales larger than the box size. In Paper I, we have demonstrated that if the initial electron temperature is less than $\sim 20\%$ of the ion temperature, the wavevector of the dominant instability is aligned with the ordered magnetic field. It follows that the evolution of the dominant mode can be conveniently captured by means of 1D simulations with the computational box oriented along $x$, which we will be employing in this work. Yet, all three components of electromagnetic fields and particle velocities are tracked. 

The focus of this work is to assess how the efficiency of electron heating depends on the properties of the flow. We vary the ion plasma beta
\be
\beta_{0i}=\frac{8 \pi n_0 k_B T_{0i}}{B_0^2}~~~,
\ee
from $\beta_{0i}=5$ up to 80. Here, $n_0$ is the particle number density at the initial time and $T_{0i}$ the initial ion temperature (ions, as well as electrons, are initialized with a Maxwellian distribution). 
The electron thermal properties are specified via the electron  beta 
\be
\beta_{0e}=\frac{T_{0e}}{T_{0i}}\, \beta_{0i}~~.
\ee
 We vary the ratio $\beta_{0e}/\beta_{0i}$ from $\ex{1}$ down to $\ex{3}$.
 The magnetization is quantified by the Alfv\'en speed
\be
v_{A0i}=\frac{B_0}{\sqrt{4 \pi m_i n_0}}~~~,
\ee
so that the initial ion temperature equals $k_B T_{0i}=m_i\beta_{0i} v_{A0i}^2/2$. We will show that our results are the same when varying the Alfv\'en velocity from 0.025 up to 0.1, with $v_{A0i}/c=0.05$ being our reference choice.
 
 The ion cyclotron frequency, which is related to the ion plasma frequency by $\omega_{0ci}=(v_{A0i}/c)\, \omega_{0\rm pi}$, will set the characteristic  unit of time. In particular, we will scale the compression rate $q$ to be a fraction of the ion cyclotron frequency $\omega_{0ci}$. In accretion flows, we expect the ratio $\omega_{0 ci}/q$ to be much larger than unity. We explore the range $\omega_{0 ci}/q=50-3200$, showing that our results converge in the limit $\omega_{0 ci}/q\gg1$. We evolve the system up to a few compression timescales.
 
Since we are interested in capturing the efficiency of {\it electron} heating due to {\it ion} instabilities, we need to properly resolve the kinetic physics of both ions and electrons. We typically employ 32,768 computational particles per species per cell, but we have tested that our results are the same when using up to 131,072 particles per species per cell. Such large values are of critical importance to suppress the spurious heating induced by the coarse-grained description of PIC plasmas \citep[e.g.,][]{melzani_13}, and so to reliably estimate the efficiency of electron heating by ion velocity-space instabilities, which is the primary focus of this work. 

We resolve the electron skin depth $c/\omega_{0\rm pe}=\sqrt{c^2 m_e/4\pi n_0 e^2}\,$ with 5 cells, which is sufficient to capture the physics of the electron whistler instability \citep{kennel_66,ossakow_72,ossakow_72b,yoon_87,yoon_11}.
On the other hand, our computational domain needs to be large enough to include at least a few wavelengths of ion-driven instabilities, i.e., a few ion Larmor radii 
\be
r_{L,i}=\sqrt{\frac{3\,\beta_{0i} \,m_i}{2\,m_e}}\, \frac{c}{\omega_{0\rm pe}}~~.
\ee
Most of our results employ a reduced mass ratio ($m_i/m_e=16$ or 64), for computational convenience. However, we show that our conclusions can be readily rescaled up to realistic mass ratios (we extend our study up to $m_i/m_e=1024$). For $m_i/m_e=16$ and $\beta_{0i}=20$, which will be our reference case, we employ a 1D computational box with $L_x=1536\,{\rm cells}\sim15\,r_{L,i}$. When changing $m_i/m_e$ or $\beta_{0i}$, we ensure that our computational domain is scaled such that it contains at least $\sim 10 \,r_{L,i}$, so that the dominant wavelength of ion-driven instabilities is properly resolved.

\section{Electron Heating by the Ion Cyclotron Instability}\label{sec:heating}
In this Section, we first summarize the main conclusions of Paper I. Then, we describe how the efficiency of electron heating depends on the flow conditions.

By means of 2D simulations, we have demonstrated in Paper I that, if the initial electron-to-ion temperature ratio is $T_{0e}/T_{0i}\lesssim0.2$, the ion cyclotron instability dominates the relaxation of the ion anisotropy, over the competing mirror mode. Since the wavevector of the ion cyclotron instability is aligned with the mean magnetic field, the relevant physics can be conveniently studied by means of 1D simulations, with the box aligned with the ordered field. Thanks to the greater number of computational particles per cell allowed by 1D simulations, as opposed to 2D, the efficiency of electron heating by the ion cyclotron instability can be reliably estimated.

Since we are interested in the {\it net heating of electrons by the ion cyclotron instability, and not in the straightforward effect of compression}, we have chosen to quantify the efficiency of electron heating by defining the ratio
\be\label{eq:chi}
\chi\equiv\frac{m_i}{m_e}\frac{\la p_e^2\ra}{\la p_i^2\ra}~~,
\ee
where $p_e=\gamma_e m_e v_e$ and $p_i=\gamma_i m_i v_i$ are the electron and ion momenta. The ratio $\chi$ 
 remains constant before the growth of anisotropy-driven instabilities, since in a compressing box the particle parallel momentum does not change, whereas the perpendicular momentum increases as $p_{\perp}\propto (1+q\,t)$, so the total momentum $p=(p_\parallel^2+2\,p_{\perp}^2)^{1/2}$ increases as $p=p_0\sqrt{[1+2\,\qt^2]/3}$, for both electrons and ions. It follows that the $\chi$ parameter is a good indicator of the electron energy change occurring as a result of the  ion cyclotron mode. The choice of the pre-factor $m_i/m_e$ is such that for  non-relativistic particles, having $\la p_e^2\ra\simeq m_e^2\la v_e^2\ra$ and $\la p_i^2\ra\simeq m_i^2\la v_i^2\ra$,  the parameter $\chi$ reduces to $\chi=m_e\la v_e^2\ra/m_i\la v_i^2\ra $, i.e., to the ratio of the average kinetic energies of electrons and ions.

As described in Paper I, the development of the ion cyclotron instability causes strong electron heating. The resulting change in the electron Lorentz factor $\la\dg_{e,q}\ra$, averaged over all the electrons in the system, gives a corresponding increase in the $\chi$ parameter of
\be\label{eq:deltachi}
\Delta\chi\simeq\frac{m_e}{m_i}\frac{\la \gamma_e \dg_{e,q}\ra c^2}{\la \gamma_i^2 v_i^2\ra}~~~,
\ee
where we have neglected higher order terms in $\dg_{e,q}/\gamma_e\ll1$, as we have motivated in Paper I. In turn, the mean energy gain $\la\Delta \gamma_{e,q} m_e c^2\ra$ at the end of the exponential growth of the ion cyclotron mode can be written as a sum of different terms (see Paper I)
\be\label{eq:heat0}
\!\!\!\!\!\!\!\!\!\!\!\!\!\!\la\dg_{e,q}\ra\equiv\la\gamma_e - \gamma_{e,q}\ra&=&\la-|\Delta \gamma_{\rm curv}|+\nonumber\\&&+\Delta \gamma_{\partial B}\!+\!\Delta \gamma_{\grad B}\!+\!\Delta \gamma_{E\cross B}\ra~,
\ee
where we have subtracted $\gamma_{e,q}$ on the left hand side, since it describes the electron energy increase due to compression alone, which would be present even without any instability. The four terms on the right hand side of \eq{heat0} can be written as a function of the plasma properties  at the end of the exponential  phase of ion cyclotron growth (as indicated by the subscript ``exp'' below)
\be
\!\!\langle\Delta \gamma_{\partial B}\!-\!|\Delta \gamma_{\rm curv}|\ra\!&\simeq&\! \left[\frac{k_B (T_\perpe-T_\pare)}{m_ec^2}\,\frac{\la \delta B_\perp^2\ra}{2|\la \bvec \ra|^2}\right]_{\rm exp}\label{eq:heat1}\\
\!\!\langle\Delta \gamma_{\grad B}\ra\!&\simeq&\! \left[\frac{k_B T_\perpe}{m_ec^2}\,\frac{\la \delta B_\perp^2\ra^2}{8|\la \bvec \ra|^4}\right]_{\rm exp}\label{eq:heat2}\\
\!\!\langle\Delta \gamma_{E\times B}\ra\!&\simeq&\! \left[\frac{\la \delta E_\perp^2\ra}{2|\la \bvec \ra|^2}\right]_{\rm exp}\!\!\!\simeq\!\left[\!\left(\frac{\omega_i}{k_i c}\right)^{\!2}\!\!\frac{\la \delta B_\perp^2\ra}{2|\la \bvec \ra|^2}\right]_{\rm exp}\label{eq:heat3}
\ee
where \eq{heat1} accounts for the sum of the energy gain due to the magnetic field growth and the energy loss associated with the curvature drift; \eq{heat2} describes the effect of the grad-B drift; and \eq{heat3} accounts for the energy gain associated with the E-cross-B velocity. We refer to Section 4.1 of Paper I for a complete characterization of the different terms.

In Eqs.~\eqn{heat1}-\eqn{heat3}, $T_\perpe$ and $T_\pare$ are the electron temperatures perpendicular and parallel to the mean field $\bavg$, and we have defined the space-averaged fields $\la \delta B_{\perp}^2\ra=\la \delta B_y^2\ra+\la\delta B_z^2\ra$ and $\la \delta E_{\perp}^2\ra=\la \delta E_y^2\ra+\la\delta E_z^2\ra$. From  Maxwell's equation in \eq{basic1}, the electric field energy is related to the magnetic field energy by $\la \delta E_\perp^2\ra=(\omega_i/k_i c)^2\la \delta B_\perp^2\ra$, which we have used in \eq{heat3}. Here, $\omega_i/k_i$ is the phase speed of ion cyclotron waves. At the condition of marginal stability, the ion anisotropy approaches \citep[e.g.,][see also Paper I]{gary_76,gary_94,gary_94d,gary_97,hellinger_06,yoon_12b,yoon_13}
\be\label{eq:margi}
A_{i,\rm MS}=\left[\frac{\beta_\perpi}{\beta_{\pari}}-1\right]_{\rm MS}\simeq\frac{\coeffi}{\beta_{\pari}^{\powi}}~~~.
\ee
The oscillation frequency and the dominant wavevector of the ion cyclotron mode at marginal stability are \citep{kennel_66,davidson_75,yoon_92,yoon_10,schlickeiser_10}
\be\label{eq:ki}
k_i&=&\frac{\omega_{\rm pi}}{c}\frac{A_{i,\rm MS}}{\sqrt{A_{i,\rm MS}+1}}~~,\\
\omega_i&=&\omega_{ci}\frac{A_{i,\rm MS}}{A_{i, \rm MS}+1}\label{eq:omegai}~~,
\ee
so that the phase speed to be used in \eq{heat3} reduces to $\omega_i/k_i=v_{Ai}/\sqrt{A_{i,\rm MS}+1}$. Finally, since $A_{i,\rm MS}\ll1$ for $\beta_{0i}=5-80$, as employed in this work, the expressions above can be simplified to give $k_i\simeq A_{i,\rm MS}\, \omega_{\rm pi}/c$ and $\omega_i\simeq A_{i,\rm MS} \,\omega_{ci}$,  so that $\omega_i/k_i\simeq v_{Ai}$. Given the scaling $k_i\propto A_{i,\rm MS}\propto \beta_\pari^{-\powi}$, it follows that the ratio of the dominant wavelength $\lambda_i=2\pi/ k_i\propto \beta_\pari^\powi\, c/\omega_{\rm pi}$ to the ion Larmor radius $r_{L,i}\propto\beta_{0i}^{1/2}\,c/\omega_{\rm pi}$ should be nearly a constant, as we indeed confirm in \sect{beta}.

In order to compute the various heating terms in Eqs.~\eqn{heat1}-\eqn{heat3}, we  need an estimate of the ratios $\la \delta B_\perp^2\ra/|\bavg|^2$ and $\la \delta E_\perp^2\ra/|\bavg|^2$ at the end of the exponential phase of the ion cyclotron instability. Moreover, for \eq{heat1} we will need the value of $T_{\perpe}/T_\pare-1$ at the saturation of the instability. All the other ingredients (e.g., $T_\perpe$ in \eq{heat2} and $\omega_i/k_i\simeq v_{Ai}$ in \eq{heat3}) can be estimated from their initial values, using the scalings expected as a result of compression alone.

As we demonstrate in \sect{flow} and justify analytically in \app{sat}, the following scalings can be employed to fully characterize the efficiency of electron heating by the ion cyclotron instability:
\be
\left[\frac{\la \delta B_{\perp}^2\ra}{|\bavg|^2}\right]_{\rm exp}\!\!&\sim\,&0.3 \left(\frac{\beta_{0i}}{20}\right)^{0.5}\label{eq:cond1}\\
\left[\frac{\la \delta E_{\perp}^2\ra}{|\bavg|^2}\right]_{\rm exp}\!\!&\sim\,&\left[\frac{\la \delta B_{\perp}^2\ra}{|\bavg|^2}\frac{v_{Ai}^2}{c^2}\right]_{\rm exp}\frac{1}{A_{i,\rm MS}+1}\label{eq:cond2}\\
\left[\frac{T_\perpe}{T_\pare}-1\right]_{\rm exp}\!\!&\sim\,\,&{\rm min}\,[A_{e,\rm MS},\,5\,A_{i,\rm MS}]\label{eq:cond3}
\ee
where $v_{Ai}=v_{A0i}\qt$. The anisotropy $A_{i,\rm MS}$ should be evaluated at the threshold of marginal stability for the ion cyclotron mode, following \eq{margi}. In the same way, the electron anisotropy $A_{e,\rm MS}$ in \eq{cond3} is to be computed at the threshold of marginal stability for the electron whistler instability, given by  \citep[][see also Paper I]{gary_96,gary_06}
\be\label{eq:marge}
A_{e,\rm MS}=\left[\frac{\beta_\perpe}{\beta_{\pare}}-1\right]_{\rm MS}\simeq\frac{\coeffe}{\beta_{\pare}^{\powe}}~~,
\ee
 The coefficients in \eq{cond1} and \eq{cond3} have been fitted using simulations with $\omega_{0ci}/q=100$. As we argue in \sect{mag}, in the astrophysically-relevant limit $\omega_{0ci}/q\gg1$, the coefficients asymptote to a value that is just a factor of a few smaller.

The scalings in Eqs.~\eqn{cond1}-\eqn{cond3} --- which will be extensively tested in the following sections --- provide, together with Eqs.~\eqn{heat1}-\eqn{heat3}, a complete characterization of the efficiency of electron heating by the ion cyclotron instability in accretion flows. In particular, \eq{cond3} shall be used to estimate the electron anisotropy that enters \eq{heat1}. Using Eqs.~\eqn{cond1}-\eqn{cond3}, we shall now estimate how the different contributions to electron heating in Eqs.~\eqn{heat1}-\eqn{heat3} depend on the flow conditions. 

We assume that the exponential growth of the ion cyclotron instability terminates at $q\,t\simeq0.5$, which is supported by our findings in \sect{flow}. The plasma conditions at this time can be estimated from their initial values, using the scalings resulting from compression (e.g., $v_{Ai}=v_{A0i}\qt$; also, $\beta_\pari=\beta_{0i} \qt^{-2}$ and $\beta_\perpi=\beta_{0i}$ for non-relativistic ions). This allows us to emphasize how the various contributions to electron heating depend on the initial flow properties:
\be
\!\!\!\!\!\!\!\!\!\!\!\!\!\!\!\!\frac{\langle\Delta \gamma_{\partial B}\!-\!|\Delta \gamma_{\rm curv}|\ra m_e c^2}{k_B T_{0i}}&\simeq&0.03\, \frac{T_{0e}}{T_{0i}}{\rm min}\!\!\left[\frac{\beta_{0i}^{0.5}}{\beta_{0e}^{\powe}},6\,\beta_{0i}^{0.1}\right]\label{eq:scal1}\\
\!\!\!\!\!\!\!\!\!\!\!\!\!\!\!\!\frac{\langle\Delta \gamma_{\grad B}\ra m_e c^2}{k_B T_{0i}}&\simeq&0.001\,\frac{T_{0e}}{T_{0i}}\beta_{0i}\label{eq:scal2}\\
\!\!\!\!\!\!\!\!\!\!\!\!\!\!\!\!\frac{\langle\Delta \gamma_{E\times B}\ra m_e c^2}{k_B T_{0i}}&\simeq&0.15\,\frac{m_e}{m_i}\beta_{0i}^{-0.5}\frac{1}{1+0.90/\beta_{0i}^\powi}\label{eq:scal3}
\ee
Eqs.~\eqn{scal1}-\eqn{scal3}, together with $\la\Delta \gamma_{e,q}\ra$ from  \eq{heat0}, fully characterize the efficiency of electron heating due to the ion cyclotron instability in collisionless accretion flows.

From these equations, the resulting increase in the $\chi$ parameter of \eq{deltachi} can be easily obtained. For the sake of simplicity, let us now neglect any explicit time dependence (yet, we include all the appropriate powers of $\qt$ in the numerical estimates of $\Delta \chi$ that we present in the following sections). With this approximation, we find that $m_i \la \gamma_i^2 v_i^2\ra\sim m_i\la \gamma_i v_i^2\ra\simeq 3\,k_B T_{0i}$, since the ions are non-relativistic. Also, in the numerator of \eq{deltachi}, we can approximate $\la \gamma_e \ra\sim \la \gamma_{0e} \ra$, where $\la \gamma_{0e}\ra\!\simeq\! (3k_BT_{0e}/2m_e c^2)\!+\![(3k_BT_{0e}/2m_e c^2)^2+1]^{1/2}$ is the mean electron Lorentz factor at the initial time, which reduces to $\la \gamma_{0e}\ra\simeq1$ for $k_BT_{0e}/m_e c^2\ll1$ (non-relativistic electrons) and to $\la \gamma_{0e}\ra\simeq 3\,k_BT_{0e}/m_e c^2$ for $k_BT_{0e}/m_e c^2\gg1$ (ultra-relativistic electrons). The increase in the $\chi$ parameter
\be\label{eq:delchieff}
\Delta \chi\simeq \frac{ \la \gamma_{0e}  \ra \la\Delta \gamma_{e,q} m_e c^2\ra}{3\,k_B T_{0i}}
\ee
can then be computed for a range of flow conditions. 

\section{Dependence on the Flow Conditions}\label{sec:flow}
In this section, we discuss how the efficiency of electron heating depends on the physical conditions of the flow. We explore the role of the electron-to-ion temperature ratio in \sect{tratio}, of the ion plasma beta in \sect{beta}, of the Alfv\'en velocity in \sect{sigma}, of the compression timescale (in units of $\omega_{0ci}^{-1}$) in \sect{mag} and of the ion-to-electron mass ratio in \sect{mime}. In doing so, we will confirm the scalings anticipated in Eqs.~\eqn{cond1}-\eqn{cond3}, which are at the basis of our heating model.

\begin{figure*}[!tbp]
\begin{center}
\includegraphics[width=1\textwidth]{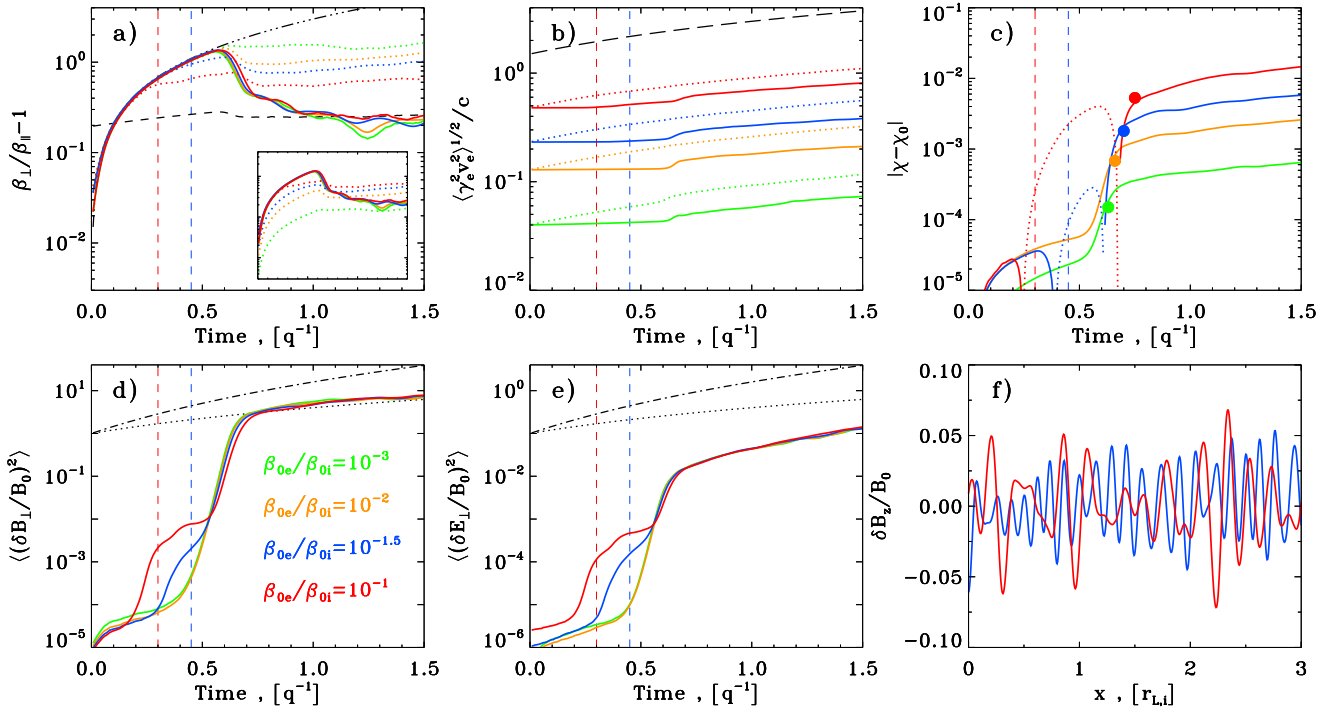}
\caption{Temporal and spatial development of compression-driven instabilities, for different choices of the electron-to-ion temperature ratio $T_{0e}/T_{0i}=\beta_{0e}/\beta_{0i}$, as detailed in the legend of panel (d). We fix $\beta_{0i}=20$, $v_{A0i}/c=0.05$, $\omega_{0ci}/q=50$ and $m_i/m_e=64$. We show the temporal evolution of the following quantities, with $q^{-1}$ as our time unit ($q$ being the compression rate): 
(a) Ion (solid) and electron (dotted) anisotropy $\beta_\perp/\beta_\parallel-1$, together with the threshold at marginal stability for the ion cyclotron instability (dashed black line, see \eq{margi}). The triple-dot-dashed black line follows the track expected from compression alone. In the subpanel, the electron anisotropy (dotted lines) is multiplied by $\beta_{0e}^\powe$, to emphasize the dependence $A_{e,\rm MS}\propto \beta_\pare^{-\powe}$ at the threshold of marginal stability for the whistler mode.
(b) Mean momentum dispersion of electrons, along (solid) or perpendicular (dotted) to the mean field. The dashed black line shows the expected evolution of the transverse component, due to compression alone (i.e., $\propto \qt$).
(c) Efficiency of electron heating or cooling, as quantified by the $\chi$ parameter defined in \eq{chi}  relative to its initial value $\chi_0$ (solid if $\chi-\chi_0\ge0$, dotted if $<0$). The filled circles show the predictions of our analytical model  at the end of the exponential growth of the ion cyclotron instability (see \sect{heating}).
(d) Magnetic energy density in the transverse component $\la \delta B_\perp^2\ra=\la \delta B_y^2+\delta B_z^2\ra$, in units of $B_0^2$. For reference, we also plot $\qt^2$ (dotted black line) and $\qt^4$ (dot-dashed black line, as expected for the evolution of the mean field energy). 
(e) Electric energy density in  the transverse component $\la \delta E_\perp^2\ra=\la \delta E_y^2+\delta E_z^2\ra$, in units of $B_0^2$. The dotted and the dot-dashed black lines are the same as in panel (d), apart from the normalization.
Finally, panel (f) presents --- at the times indicated with  the vertical dashed lines in all the other panels --- the spatial profile of $\delta B_z/B_0$, as a function of the longitudinal coordinate $x$, measured in units of the ion Larmor radius $r_{L,i}$. }
\label{fig:tratio}
\end{center}
\end{figure*}

\begin{figure*}[!htbp]
\begin{center}
\includegraphics[width=1\textwidth]{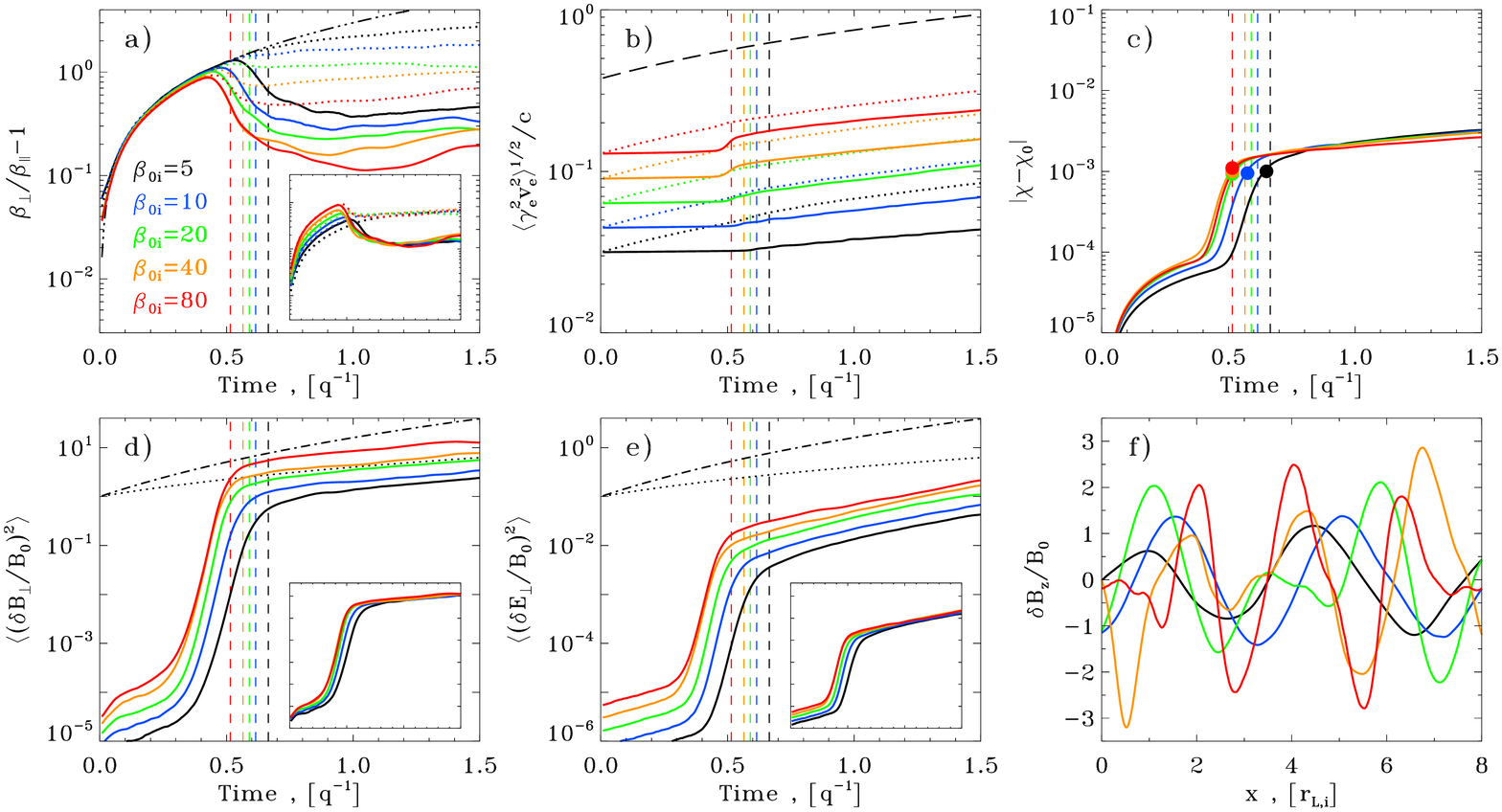}
\caption{Temporal and spatial development of compression-driven instabilities, for different choices of the ion plasma beta $\beta_{0i}$, as detailed in the legend of panel (a). We fix $\beta_{0e}/\beta_{0i}=\ex{2}$, $v_{A0i}/c=0.05$, $\omega_{0ci}/q=100$ and $m_i/m_e=16$. See the caption of \fig{tratio} for details. In the subpanel of \fig{beta}(a), we multiply the anisotropy of both ions and electrons by $ \beta_{0i}^{\powi}$,  showing that this leaves no residual dependence at late times, when the ion anisotropy approaches the threshold of marginal stability in \eq{margi}. In the subpanels of \fig{beta}(d) and (e), the magnetic and electric energy densities are multiplied by $\beta_{0i}^{-0.5}$, showing that this leaves no residual dependence (see \app{sat} for an analytical justification of this scaling). }
\label{fig:beta}
\end{center}
\end{figure*}

In Figs.~\fign{tratio}--\fign{mimehigh}, we present the temporal evolution of our system, for different initial conditions. In each case, we show in the top row the time evolution of the particle anisotropy (panel (a); solid lines for ions, dotted for electrons), of the electron momentum (panel (b); solid for the parallel component, dotted for the perpendicular component) and of the electron heating efficiency (panel (c)), quantified by  $|\chi-\chi_0|$, where the parameter $\chi$ is defined in \eq{chi} and $\chi_0$ is its value  at the initial time. In the bottom row, we show the time evolution of the trasverse magnetic energy (panel (d)) and of the transverse electric energy (panel (e)). In addition, we show in panel (f) the spatial pattern (with $r_{L,i}$ as our unit of length) of the $\delta B_z$ component of the unstable waves.

Below, we focus on a few representative cases, but we remark that the conclusions presented below have been checked across the whole parameter space explored in this work, i.e., for $\beta_{0i}=5-80$, for $\beta_{0e}/\beta_{0i}=\ex{3}-\ex{1}$, for $v_{A0i}=0.025-0.1$, for $\omega_{0ci}/q=50-3200$ and up to a mass ratio $m_i/m_e=1024$.

\subsection{Dependence on the Temperature Ratio}\label{sec:tratio}
In this section, we illustrate the dependence on the electron-to-ion temperature ratio $T_{0e}/T_{0i}=\beta_{0e}/\beta_{0i}$, for a representative case with  $\beta_{0i}=20$, $v_{A0i}/c=0.05$, $\omega_{0ci}/q=50$ and $m_i/m_e=64$. We vary the electron-to-ion temperature ratio from $\beta_{0e}/\beta_{0i}=\ex{3}$ up to $\ex{1}$. 

In the range $T_{0e}/T_{0i}\lesssim 0.1$ explored in \fig{tratio}, the development of the ion cyclotron instability is not sensitive to the electron thermal content. In fact, \fig{tratio} shows that the time of growth of the ion cyclotron instability ($q\,t\simeq 0.6$) and the saturation values of the magnetic and electric energy ($\la\delta B_\perp^2\ra/|\bavg|^2\sim0.3$ in \fig{tratio}(d) and $\la\delta E_\perp^2\ra/|\bavg|^2\sim 3\times \ex{3}$ in \fig{tratio}(e), respectively) are all insensitive to the electron-to-ion temperature ratio. Moreover, the secular evolution that follows the exponential phase of the ion cyclotron instability (at $q\,t\gtrsim 0.7$) is the same, regardless of $T_{0e}/T_{0i}$. Once the energy of the ion cyclotron waves reaches $\la \delta B_\perp^2\ra^{1/2}/|\bavg|\sim0.1$, efficient pitch-angle scattering brings the ion anisotropy (\fig{tratio}(a)) back to the threshold of marginal stability in \eq{margi} (which is indicated in \fig{tratio}(a) with a dashed black line). At $q\,t\gtrsim 0.9$, the ion anisotropy follows the same track of marginal stability, regardless of the initial electron temperature. In short, the ion physics does not depend on the electron thermal content.

The choice of initial electron temperature can affect the electron physics before the onset of the ion cyclotron instability. For cold electrons ($T_{0e}/T_{0i}=\ex{3}$ in green and $T_{0e}/T_{0i}=\ex{2}$ in orange), the ion cyclotron instability starts to grow when the electron anisotropy has not reached yet the threshold for the electron whistler instability in \eq{marge} (corresponding to $A_{e,\rm MS}\simeq 1.3 $ for $T_{0e}/T_{0i}=\ex{2}$ and to $A_{e,\rm MS}\simeq 3.9 $ for $T_{0e}/T_{0i}=\ex{3}$). In this case, the electron anisotropy is not strong enough to trigger anisotropy-driven instabilities on electron scales, and no signs of electron physics are present. In particular, the temporal evolution of the magnetic and electric energy (\fig{tratio}(d) and (e), respectively) is the same at all times, regardless of the electron temperature. The only dependence on $T_{0e}/T_{0i}$ in this regime of cold electrons ($T_{0e}/T_{0i}\lesssim\ex{2}$) appears in the value of electron anisotropy after the growth of the ion cyclotron instability (compare the green and orange dotted lines in \fig{tratio}(a) at $q\,t\gtrsim 0.6$). The degree of electron anisotropy is regulated by the mechanism of electron heating during the ion phase, as we detail below.\footnote{As we have described in Paper I (Section 4.2.1), the electron distribution for $\beta_{0e}\lesssim 2\, m_e/m_i$ is not uniform in space, due to the local nature of the heating process (which is dominated by the term in \eq{heat3}). The electron distribution resembles a Maxwellian drifting with the local E-cross-B velocity, so the degree of electron anisotropy shown by the green dotted line in \fig{tratio} is a measure of the E-cross-B speed, rather than of the genuine electron anisotropy in the fluid frame.}

At higher electron temperatures ($T_{0e}/T_{0i}=\ex{1.5}$ in blue and $T_{0e}/T_{0i}=\ex{1}$ in red), the threshold for the whistler instability --- which corresponds to $A_{e,\rm MS}\simeq 0.7 $ for $T_{0e}/T_{0i}=\ex{1.5}$ and to $A_{e,\rm MS}\simeq 0.4 $ for $T_{0e}/T_{0i}=\ex{1}$ --- is reached before the onset of the ion cyclotron mode. So, the whistler instability can grow, generating transverse electromagnetic waves (see Appendix B in Paper I for further details), and producing a bump at early times ($q\,t\simeq 0.4$)  in the magnetic and electric energy of \fig{tratio}(d) and (e), respectively.\footnote{As discussed in Paper I, our 2D simulations show that the wavevector of the fastest growing whistler mode is aligned with the mean magnetic field, so 1D simulations can properly capture the development of the whistler instability.}  In \app{sat}, we estimate analytically the saturation values of the magnetic and electric energy produced by the whistler instability, as a function of the flow conditions. Our estimates are in good agreement with the results of our simulations, regarding the dependence on both $T_{0e}/T_{0i}$ (investigated in \fig{tratio}) and $m_i/m_e$ (see \fig{mimehigh} for details).

During the whistler phase at $0.3\lesssim q\,t \lesssim0.55$, the electron anisotropy remains at the threshold of marginal stability, see the dotted blue and red lines in \fig{tratio}(a). In contrast, the ion anisotropy still follows the track $\qt^2-1$ expected from compression alone (indicated by the triple-dot-dashed line in \fig{tratio}(a)), suggesting that ions do not participate in the whistler instability. We expect the electron anisotropy during the whistler phase ($0.3\lesssim q\,t \lesssim0.55$) to scale as $A_{e,\rm MS}\propto \beta_{\pare}^{-\powe}$, as in \eq{marge}. Once we multiply the electron anisotropy by a factor of $\beta_{0e}^{\powe}$, we find that the temporal tracks for $T_{0e}/T_{0i}=\ex{1.5}$ and $\ex{1}$ almost overlap (blue and red dotted lines in the subpanel of \fig{tratio}(a)), indicating that the condition of marginal stability in \eq{marge} indeed regulates the level of electron anisotropy after the onset of the whistler instability. While it is not suprising that the electron anisotropy stays at the threshold $\sim A_{e,\rm MS}$ during the electron phase at $0.3\lesssim q\,t \lesssim0.55$, we remark that a similar degree of anisotropy is preserved after the growth of the ion cyclotron instability at $q\,t\gtrsim 0.6$, which justifies our {\it ansatz} in \eq{cond3} (see the first term in the square brackets).

The electron whistler instability generates transverse electromagnetic waves on electron scales.  At marginal stability, the characteristic wavelength and oscillation frequency of the electron whistler mode are respectively \citep{kennel_66,yoon_87,yoon_11,bashir_13}
\be
\lambda_e\simeq \frac{2\pi}{A_{e,\rm MS}^{1/2}}\,\frac{c}{\omega_{\rm pe}}~~{\rm and}~~\omega_e\simeq \frac{A_{e,\rm MS}}{A_{e,\rm MS}+1}\, \omega_{ce}~.
\ee
Since we account for the possibility that electrons are ultra-relativistic, the proper definitions of the electron cyclotron frequency and plasma frequency are 
\be
\omega_{ce}=\frac{e|\bavg|}{\la\gamma_e\ra m_e c}~~{\rm and}~~\omega_{\rm pe}=\sqrt{\frac{4 \pi n e^2}{\la \gamma_e \ra m_e}}~~.
\ee
 In \fig{tratio}(f), we show the spatial pattern of the $\delta B_z$ component of whistler waves, measured at the time indicated with the blue and red dashed vertical lines in all the other panels (for $T_{0e}/T_{0i}=\ex{1.5}$ and $\ex{1}$, respectively). In units of the ion Larmor radius $r_{L,i}$, the characteristic wavelength of the whistler instability is
\be
\frac{\lambda_e}{r_{L,i}}\simeq 2\pi A_{e,\rm MS}^{-1/2}\sqrt{\frac{2 \la \gamma_e\ra m_e}{3\, \beta_{0i} \,m_i}}~~~.\label{eq:lambdae}
\ee
In \fig{tratio}, we employ $\beta_{0i}=20$ and $m_i/m_e=64$. For $T_{0e}/T_{0i}=\ex{1.5}$, taking $A_{e,\rm MS}\simeq 1$ from the dotted bue line in \fig{tratio}(a), we expect from \eq{lambdae} that $\lambda_e/r_{L,i}\simeq 0.2$, which is in good agreement with the spatial periodicity of the blue line in \fig{tratio}(f). Similarly, for $T_{0e}/T_{0i}=\ex{1}$, taking $A_{e,\rm MS}\simeq 0.6$ from the dotted red line in \fig{tratio}(a), we expect $\lambda_e/r_{L,i}\simeq 0.25$, which agrees with the spatial pattern of the red line in \fig{tratio}(f). We have extensively checked the scalings in \eq{lambdae}, most importantly regarding the dependence on $T_{0e}/T_{0i}$ and $m_e/m_i$. In the limit of non-relativistic electrons (so, $\la \gamma_e\ra\simeq1$), we find from \eq{lambdae} that $\lambda_e/r_{L,i}\propto \beta_{\pare}^{0.25}$, since $A_{e,\rm MS}\propto \beta_{\pare}^{-\powe}$. This scaling is consistent with the numerical findings of \citet{gary_06}.

We conclude by discussing the efficiency of electron heating, as a function of the electron-to-ion temperature ratio. In \fig{tratio}(c), we plot the temporal evolution of the $\chi$ parameter defined in \eq{chi} relative to its initial value $\chi_0$ (solid lines if $\chi-\chi_0\ge0$, dotted if $<0$). The filled circles in \fig{tratio} show, for different choices of $T_{0e}/T_{0i}$, our analytical prediction for the change in the $\chi$ parameter --- as derived from the theory outlined in \sect{heating} --- at the end of the exponential phase of the ion cyclotron instability. Regardless of the initial electron temperature, the growth of the ion cyclotron waves at $q\, t\simeq 0.7$ leads to substantial electron heating, whose magnitude is properly captured by our analytical model (compare the lines with the filled circles of the same color). The increase in the $\chi$ parameter is more pronounced for higher values of $T_{0e}/T_{0i}$. This can be easily understood from the scalings in Eqs.~\eqn{scal1}-\eqn{scal3}, as we now explain.

At small values of $T_{0e}/T_{0i}$ (green line in \fig{tratio}(c), for $T_{0e}/T_{0i}=\ex{3}$), electron heating by the ion cyclotron instability is dominated by the E-cross-B term in \eq{scal3}, which is insensitive to the electron temperature. In Paper I, we have demonstrated that the E-cross-B contribution dominates if $\beta_{0e}\lesssim 2\,m_e/m_i$. In this case, \eq{scal3} together with \eq{delchieff} show that the increase in the $\chi$ parameter is independent of $T_{0e}/T_{0i}$ (as long as $\la \gamma_{0e}\ra\simeq 1$), which we have confirmed by running dedicated simulations with even smaller values of the electron-to-ion temperature ratio, $T_{0e}/T_{0i}=2.5\times\ex{4}$ and $T_{0e}/T_{0i}=\ex{4}$ (not shown in \fig{tratio}).

For $\beta_{0e}\gtrsim 2\, m_e/m_i$, which corresponds to $T_{0e}/T_{0i}\gtrsim 1.5\times\ex{3}$ for the parameters adopted in \fig{tratio}, the physics of electron heating is controlled by the terms in \eq{scal1} and \eq{scal2}, with the former that typically dominates, due to its weaker dependence on $\la \delta B_\perp^2\ra/|\la \bvec \ra|^2\ll1$. While the E-cross-B term accounts for as much as $64\%$ of the mean electron energy gain for $T_{0e}/T_{0i}=\ex{3}$, its contribution falls  down to $15\%$ for 
$T_{0e}/T_{0i}=\ex{2}$, and it is even smaller for higher values of $T_{0e}/T_{0i}$ (see the rightmost column in \tab{tratio}). In parallel, the combination of $\la\Delta \gamma_{\partial B}\ra>0$ and $\la\Delta \gamma_{\rm curv}\ra<0$ results in a fractional contribution to electron heating that increases from $31\%$ at  $T_{0e}/T_{0i}=\ex{3}$ up to $72\%$ at $T_{0e}/T_{0i}=\ex{2}$ and is even larger for higher $T_{0e}/T_{0i}$ (see the leftmost column in \tab{tratio}). If the term in \eq{scal1} dominates over the contribution in \eq{scal2}, as it happens for all the values of $T_{0e}/T_{0i}$ investigated in \fig{tratio} (compare the first and second column in \tab{tratio}), we expect the electron fractional energy gain to increase initially as $\propto T_{0e}/T_{0i}$, as long as the second term in the square brackets of \eq{scal1} is smaller than the first one, and then as $\propto T_{0e}^{0.5}/T_{0i}^{0.5}$. While the former scaling is the same as in the grad-B term of \eq{scal2}, the latter is shallower, which explains why for high values of $T_{0e}/T_{0i}$ the fractional contribution to electron heating by the grad-B term  becomes increasingly larger, at the expense of the term in \eq{scal1} (compare first and second columns in \tab{tratio} between $T_{0e}/T_{0i}=\ex{1.5}$ and $\ex{1}$).
 
Finally, we point out that our model tends to overestimate the actual increase in the $\chi$ parameter observed for high values of $T_{0e}/T_{0i}$ (compare the red line and the filled red circle in \fig{tratio}(c)). As we discuss in \app{cool}, this is due to the the early decrease in the $\chi$ parameter that accompanies the growth of electron whistler waves. The value of $\chi-\chi_0$ becomes as low as $\simeq -4\times\ex{3}$ at $q\,t\simeq 0.6$ (dotted red line in \fig{tratio}(c)), so that the subsequent increase driven by the ion cyclotron instability falls short of our analytical prediction, which does not take into account the degree of electron cooling during the whistler phase. Extrapolating this argument to even higher values of the electron-to-ion temperature ratio, one might be tempted to infer an upper limit on $T_{0e}/T_{0i}$, where cooling by the whistler instability balances heating by the ion cyclotron mode. However, the following caveats should be considered.

As we demonstrate in \sect{mag}, as $\omega_{0ci}/q$ increases, the ion cyclotron instability appears earlier, and in the astrophysically-relevant limit  $\omega_{0ci}/q\gg1$ it will grow soon after the ion anisotropy exceeds the threshold in \eq{margi}. Then, the condition $\coeffe/\beta_{\pare}^{\powe}\lesssim \coeffi/\beta_{\pari}^{\powi}$ required for the whistler instability to have a smaller threshold than the ion cyclotron mode, such that to precede the ion cyclotron growth, can be recast as a lower limit on the electron-to-ion temperature ratio 
\be
\frac{T_{0e}}{T_{0i}}\gtrsim \frac{0.7}{\beta_{0i}^{0.2}}\sim0.3-0.5~~~,
\ee
for the range of $\beta_{0i}\sim 5-30$ expected in accretion flows. This is already in the regime where oblique mirror modes cannot be neglected in the evolution of the system (see Paper I), so one needs to perform 2D simulations. A detailed study of electron heating in the regime $T_{0e}/T_{0i}\gtrsim 0.3$ --- which must be performed with 2D simulations having $\omega_{0ci}/q\gg1$ --- will be presented elsewhere.

\begin{table}
\centering
\caption{Fractional Contributions to Electron Heating}\label{tab:tratio}
\begin{tabular}{cccc}\hline\hline
Run & $\frac{\la\Delta\gamma_{\partial B}-|\Delta\gamma_{\rm curv}|\ra}{\la \Delta\gamma_{e,q}\ra}$ 
 & $\frac{\la\Delta\gamma_{\grad B}\ra}{\la \Delta\gamma_{e,q}\ra}$ 
 & $\frac{\la\Delta\gamma_{E\cross B}\ra}{\la \Delta\gamma_{e,q}\ra}$\\[4pt]
\hline\hline
$\beta_{0e}/\beta_{0i}=\ex{3}$ & 0.31 & 0.05 & 0.64 \\\hline
$\beta_{0e}/\beta_{0i}=\ex{2}$ &0.72 & 0.13 &  0.15 \\\hline
$\beta_{0e}/\beta_{0i}=\ex{1.5}$ &0.79 & 0.15 & 0.06 \\\hline
$\beta_{0e}/\beta_{0i}=\ex{1}$ &0.78 &  0.19 & 0.03 \\\hline\hline
\multicolumn{4}{l}{%
  \begin{minipage}{8cm}%
    Note: We fix $\beta_{0i}=20$, $v_{A0i}/c=0.05$, $\omega_{0ci}/q=50$ and $m_i/m_e=64$.%
  \end{minipage}%
}\\
\end{tabular}
\end{table}
\vspace{0.35in}

\subsection{Dependence on the Ion Plasma Beta}\label{sec:beta}
In this section, we examine the dependence of our results on the ion plasma beta $\beta_{0i}$, for a representative case with fixed $\beta_{0e}/\beta_{0i}=\ex{2}$, $v_{A0i}/c=0.05$, $\omega_{0ci}/q=100$ and $m_i/m_e=16$. We explore the range $\beta_{0i}=5-80$, but we argue that our results apply down to $\beta_{0i}\gtrsim 1$.

In \fig{beta}, we show that the level of magnetic and electric energy (panel (d) and (e), respectively) resulting from the ion cyclotron instability increases monotonically with $\beta_{0i}$ (from black to red as $\beta_{0i}$ varies from 5 up to 80). Once normalized to the mean field energy $|\la \bvec\ra|^2/8\pi$ --- whose temporal evolution is shown with a dot-dashed black line in \fig{beta}(d) --- we find that for $\beta_{0i}=20$  (green line in \fig{beta}(d)), the magnetic energy in ion cyclotron waves reaches $\la \delta B_\perp^2\ra/|\la \bvec \ra|^2\sim 0.3$ at the end of the exponential growth of the ion cyclotron mode ($q\, t\simeq 0.6$). Also, both the magnetic and the electric energies scale  as $\propto \beta_{0i}^{0.5}$. This result, which is derived analytically in \app{sat}, is confirmed in the subpanels of \fig{beta}(d) and (e), where we plot the temporal evolution of the magnetic and electric energy density multiplied by $\beta_{0i}^{-0.5}$. The fact that the different curves overlap demonstrates that \be
\frac{\la \delta B_\perp^2\ra}{|\la \bvec \ra|^2}\propto \beta_{0i}^{0.5}~~{\rm and}~~\frac{\la \delta E_\perp^2\ra}{|\la \bvec \ra|^2}\propto \beta_{0i}^{0.5}~,
\ee 
everything else being fixed. A posteriori, this lends support to our {\it ansatz} in \eq{cond1} and \eq{cond2}. Moreover, \fig{beta}(d) shows that our assumption of $\la \delta B_\perp^2\ra/|\la \bvec \ra|^2\lesssim 1$ --- which we have employed in Paper I and in \sect{heating} --- is satisfied in the range of  $\beta_{0i}$ investigated in this work.

We remark that both the normalization and the $\beta_{0i}-$dependence of \eq{cond1} are in agreement with earlier studies of the compression-driven ion cyclotron instability performed with hybrid codes \citep[e.g.,][]{hellinger_05}. The scaling  $\la \delta B_\perp^2\ra\propto\beta_{0i}^{0.5}$ had previously been found also via  hybrid simulations of undriven systems  by \citet{gary_97,gary_00}. In addition, these studies had assessed the dependence on $\beta_{0i}$ of the ion anisotropy at marginal stability, see \eq{margi}. In \fig{beta}(a), we confirm that, due to pitch-angle scattering by the transverse ion cyclotron waves, the ion anisotropy at $q\,t \gtrsim 0.6$ is reduced back to its value at the threshold of marginal stability. In agreement with \eq{margi}, the residual degree of anisotropy is smaller for higher $\beta_{0i}$.\footnote{Since the critical anisotropy threshold in \eq{margi} is smaller for higher $\beta_{0i}$, the onset of the ion cyclotron instability will occur at earlier times for higher $\beta_{0i}$, as shown in \fig{beta}.} More precisely, we confirm the scaling $A_{i,\rm MS}\propto \beta_{0i}^{-\powi}$ in the subpanel of \fig{beta}(a), where we multiply the different curves by a factor of $\propto \beta_{0i}^{\powi}$. The fact that the different solid lines overlap confirms the scaling in \eq{margi} for the {\it ion} anisotropy at marginal stability. Interestingly, the subpanel of \fig{beta}(a) shows that the {\it electron} anisotropy also scales as $\propto\beta_{0i}^{-\powi}$, and it is consistently a factor of $\sim 5$ larger than the ion anisotropy at marginal stability $A_{i,\rm MS}$ (see the dotted lines in the subpanel of \fig{beta}(a), where all the  curves have been multiplied by a factor of $\propto \beta_{0i}^{\powi}$). This justifies our {\it ansatz} in \eq{cond3}, as a constraint on the degree of electron anisotropy in the absence of the electron whistler phase (see the second term in the square brackets of \eq{cond3}).

From the ion anisotropy at marginal stability in \eq{margi}, we can write the dominant wavelength of the ion cyclotron mode as $\lambda_i\simeq 2 \pi \,A_{i,\rm MS}^{-1}\,c/\omega_{\rm pi}$, where we have taken the limit of weak anisotropy $A_{i,\rm MS}\ll1$ appropriate for $\beta_{0i}\gtrsim 1$ (see \eq{ki}). Since $A_{i,\rm MS}\propto \beta_{\pari}^{-\powi}$, we find that $\lambda_i\propto \beta_{\pari}^{\powi}c/\omega_{\rm pi}$, that is remarkably similar to the 
scaling of the ion Larmor radius: $r_{L,i}\propto \beta_{0i}^{1/2}c/\omega_{\rm pi}$. It follows that the wavelength of the dominant mode should be nearly independent of $\beta_{0i}$, once measured in units of the ion Larmor radius. More precisely, $\lambda_i\propto \beta_{0i}^{-0.1}r_{L,i}$. The dependence of the dominant wavelength on $\beta_{0i}$ is shown in \fig{beta}(f), where we plot the spatial pattern of the $\delta B_z$ component of the ion cyclotron waves, at the times indicated with the vertical dashed lines in the other panels. We confirm that the dominant wavelength is nearly the same, in units of the ion Larmor radius, with a residual tendency for shorter wavelengths at higher $\beta_{0i}$, in agreement with the expected scaling $\lambda_i\propto \beta_{0i}^{-0.1}r_{L,i}$. Similarly, we have verified that the oscillation period of the ion cyclotron waves agrees with \eq{omegai}.

Regarding electron heating by the ion cyclotron instability, \fig{beta}(c) shows that the increase in the $\chi$ parameter is nearly independent of $\beta_{0i}$. This might appear surprising, given the strong dependence on $\beta_{0i}$ of the $\la \Delta \gamma_{\grad B}\ra$ and $\la \Delta \gamma_{E\cross B} \ra$ terms in \eq{scal2} and \eq{scal3}, respectively. However, for the parameters employed in \fig{beta}, we find that if $\beta_{0i}\gtrsim 20$ the combination $\la \Delta \gamma_{\partial B}-|\Delta \gamma_{\rm curv}|\ra$ in \eq{scal1} dominates the process   of electron heating (see the first column in \tab{beta}). Since the dependence of \eq{scal1} on $\beta_{0i}$ is extremely weak ($\propto \beta_{0i}^{0.1}$, in the regime relevant for \fig{beta}), it is not surprising that the heating efficiency is independent of the ion plasma beta, at $\beta_{0i}\gtrsim 20$. At smaller $\beta_{0i}$, the fractional contribution of the term in \eq{scal1} gets smaller, in favor of the E-cross-B term in \eq{scal3} (compare the first and the last columns in \tab{beta}). Despite the change in the dominant heating channel, the efficiency of electron heating stays remarkably constant even for $\beta_{0i}\lesssim 20$, both based on our analytical model (filled circles in \fig{beta}(c)) and confirmed by our simulations (solid lines in \fig{beta}(c)). Yet, at $\beta_{0i}\lesssim20$, this apparent independence of the heating efficiency on the ion plasma beta should be regarded as a mere coincidence. At even smaller values of $\beta_{0i}$, we expect the beta-dependence of the E-cross-B term in \eq{scal3} to become apparent.



\begin{figure*}[!htbp]
\begin{center}
\includegraphics[width=1\textwidth]{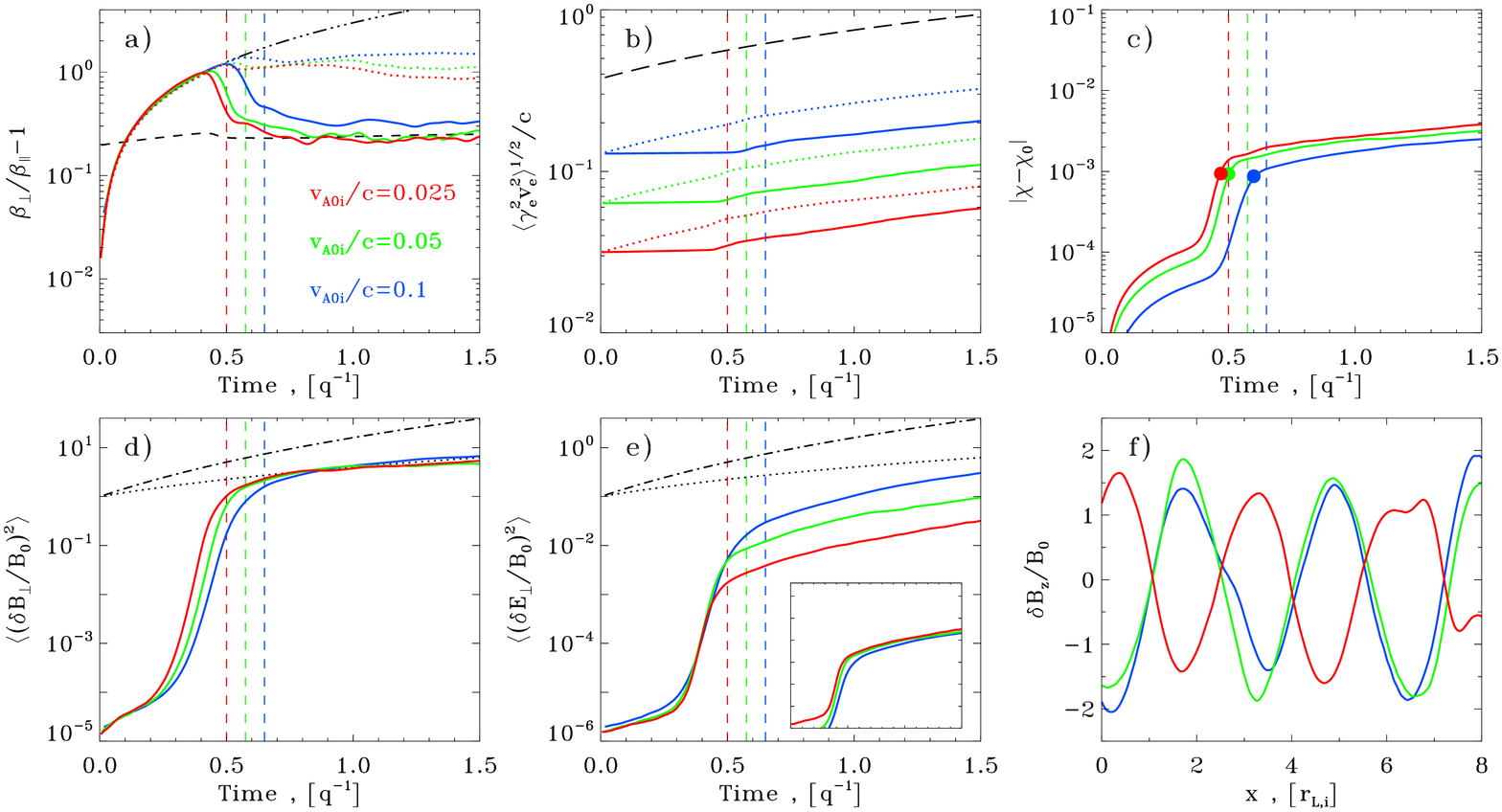}
\caption{Temporal and spatial development of compression-driven instabilities, for different choices of the Alfv\'en velocity $v_{A0i}/c$, as detailed in the legend of panel (a). We fix $\beta_{0i}=20$, $\beta_{0e}/\beta_{0i}=\ex{2}$, $\omega_{0ci}/q=100$ and $m_i/m_e=16$. See the caption of \fig{tratio} for details. In the subpanel of \fig{beta}(e), the  electric energy density is multiplied by $v_{A0i}^{-2}$, showing that this leaves no residual dependence (see \app{sat} for an analytical justification of this scaling).}
\label{fig:sigma}
\end{center}
\end{figure*}

\begin{table}
\centering
\caption{Fractional Contributions to Electron Heating}\label{tab:beta}
\begin{tabular}{cccc}\hline\hline
Run & $\frac{\la\Delta\gamma_{\partial B}-|\Delta\gamma_{\rm curv}|\ra}{\la \Delta\gamma_{e,q}\ra}$ 
 & $\frac{\la\Delta\gamma_{\grad B}\ra}{\la \Delta\gamma_{e,q}\ra}$ 
 & $\frac{\la\Delta\gamma_{E\cross B}\ra}{\la \Delta\gamma_{e,q}\ra}$\\[4pt]
\hline\hline
$\beta_{0i}=5$ & 0.30 & 0.02 & 0.68 \\\hline
$\beta_{0i}=10$ &0.40 & 0.04 &  0.56 \\\hline
$\beta_{0i}=20$ &0.49 & 0.09 & 0.42 \\\hline
$\beta_{0i}=40$ &0.56 & 0.15 & 0.29 \\\hline
$\beta_{0i}=80$ &0.56 &  0.26 & 0.18 \\\hline\hline
\multicolumn{4}{l}{%
  \begin{minipage}{8cm}%
    Note: We fix $\beta_{0e}/\beta_{0i}=\ex{2}$, $v_{A0i}/c=0.05$, $\omega_{0ci}/q=100$ and $m_i/m_e=16$.%
  \end{minipage}%
}\\
\end{tabular}
\end{table}
\vspace{0.35in}

\subsection{Dependence on the Alfv\'en Velocity}\label{sec:sigma}
In \fig{sigma} we investigate the dependence on the Alfv\'en velocity, for a representative case with fixed $\beta_{0i}=20$, $\beta_{0e}/\beta_{0i}=\ex{2}$, $\omega_{0ci}/q=100$ and $m_i/m_e=16$. We explore the range $v_{A0i}/c=0.025-0.1$.

As shown in \fig{sigma}, the temporal evolution of the magnetic energy of ion cyclotron waves (panel (d)) is nearly insensitive to the Alfv\'en velocity, in agreement with \eq{cond1} and \app{sat}. The wavelength of the ion cyclotron mode in panel (f) does not explicitly depend on $v_{A0i}/c$, at fixed $\beta_{0i}$ (see \eq{ki} combined with \eq{margi}). In addition, the temporal evolution of the ion anisotropy in \fig{sigma}(a) (solid lines) is nearly independent of $v_{A0i}/c$, in the regime of non-relativistic ions (i.e., with the marginal exception of the case $v_{A0i}/c=0.1$, where the ions are trans-relativistic). The electron anisotropy during the growth of the ion cyclotron mode is $\sim 5\, A_{i,\rm MS}$ (dotted lines in \fig{sigma}(a)), regardless of the Alfv\'en velocity, which is consistent with our {\it ansatz} in \eq{cond3} (see the second term in the square brackets).

The only quantity that depends explicitly on the Alfv\'en velocity is the electric energy density in \fig{sigma}(e), which is expected on analytical grounds to scale as $\propto v_{A0i}^2$ (see \app{sat}). We confirm this scaling in the subpanel of \fig{sigma}(e), where we multiply the different curves by $v_{A0i}^{-2}$, showing that such rescaling leaves no residual dependence on the Alfv\'en velocity. Our findings justify the dependence of \eq{cond2} on the Alfv\'en speed.

Finally, from Eqs.~\eqn{scal1}-\eqn{scal3}, we see that the electron energy change associated with the growth of the ion cyclotron mode has no explicit dependence on the Alfv\'en velocity, as confirmed in \fig{sigma}(c). The same holds for the increase in the $\chi$ parameter in \eq{delchieff}. However, we point out that this conclusion only applies to the case of non-relativistic electrons, i.e., $\la \gamma_{0e}\ra\simeq1$. If the electrons are ultra-relativistic, at fixed $\beta_{0e}$ and $m_e/m_i$ we will have that $\la \gamma_{0e}\ra\propto v_{A0i}^2$, so the electron heating efficiency as quantified by the $\chi$ parameter will increase as  $\propto v_{A0i}^2$. Albeit not shown in \fig{sigma}, we have verified this scaling in our simulations, in the parameter regime where electrons become ultra-relativistic.

\begin{table}
\centering
\caption{Fractional Contributions to Electron Heating}\label{tab:sigma}
\begin{tabular}{cccc}\hline\hline
Run & $\frac{\la\Delta\gamma_{\partial B}-|\Delta\gamma_{\rm curv}|\ra}{\la \Delta\gamma_{e,q}\ra}$ 
 & $\frac{\la\Delta\gamma_{\grad B}\ra}{\la \Delta\gamma_{e,q}\ra}$ 
 & $\frac{\la\Delta\gamma_{E\cross B}\ra}{\la \Delta\gamma_{e,q}\ra}$\\[4pt]
\hline\hline
$v_{A0i}/c=0.025$ & 0.49 & 0.09 & 0.42 \\\hline
$v_{A0i}/c=0.05$ &0.49 & 0.09 &  0.42 \\\hline
$v_{A0i}/c=0.1$ &0.49 & 0.09 & 0.42 \\\hline\hline
\multicolumn{4}{l}{%
  \begin{minipage}{8cm}%
    Note: We fix $\beta_{0i}=20$, $\beta_{0e}/\beta_{0i}=\ex{2}$, $\omega_{0ci}/q=100$ and $m_i/m_e=16$.%
  \end{minipage}%
}\\
\end{tabular}
\end{table}
\vspace{0.35in}

\begin{figure*}[tbp]
\begin{center}
\includegraphics[width=1\textwidth]{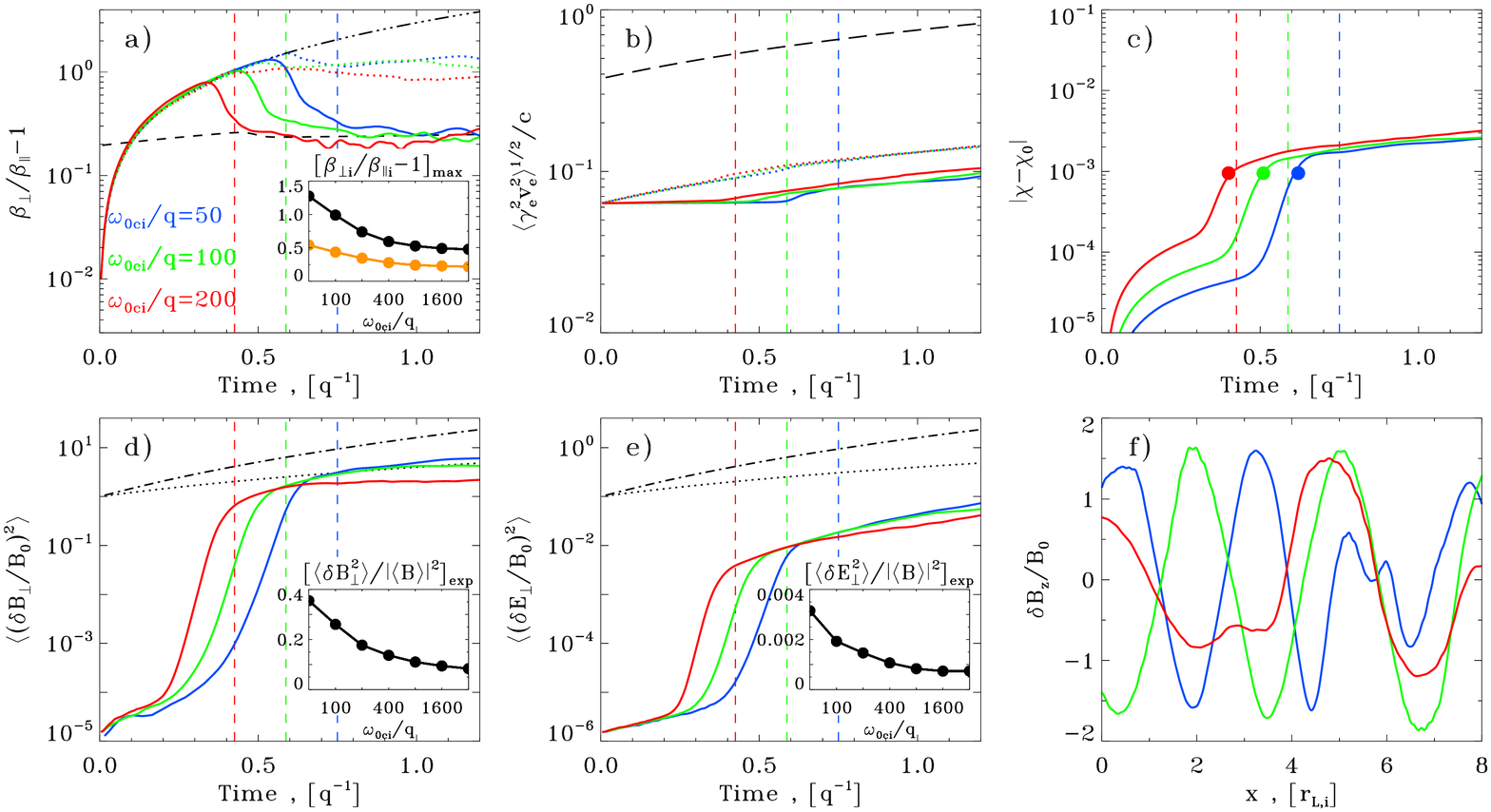}
\caption{Temporal and spatial development of compression-driven instabilities, for different choices of the parameter $\omega_{0ci}/q$. We fix $\beta_{0i}=20$, $\beta_{0e}/\beta_{0i}=\ex{2}$, $v_{A0i}/c=0.05$ and $m_i/m_e=16$, as detailed in the legend of panel (a).
In the main plot,  we examine the cases $\omega_{0ci}/q=50$ (blue lines), $\omega_{0ci}/q=100$ (green lines) and $\omega_{0ci}/q=200$ (red lines). In the subpanels of \fig{mag}(a), (d) and (e) we extend our study to $\omega_{0ci}/q=3200$. In the subpanel of \fig{mag}(a), we show both the peak value of the ion anisotropy (filled black circles) and the time at which the ion anisotropy peaks (filled orange circles). In the subpanels of \fig{mag}(d) and (e), we show respectively the magnetic and electric energies in ion cyclotron waves normalized to the mean field energy, measured at the end of the exponential phase of the ion cyclotron growth.  We remark that the horizontal axis in the subpanels of \fig{mag}(a), (d) and (e) shows the ratio $\omega_{0ci}/q$ on a logarithmic scale, from $\omega_{0ci}/q=50$ up to 3200 (with each tick mark corresponding to an increase by a factor of two). See \fig{tratio} for details.}
\label{fig:mag}
\end{center}
\end{figure*}

\vspace{-0.15in}
\subsection{Dependence on the Compression Rate}\label{sec:mag}
In this section, we analyze the dependence of our results on the compression rate, or more specifically on the ratio $\omega_{0ci}/q$, which is much larger than unity in astrophysical accretion flows. We focus on a representative case with fixed $\beta_{0i}=20$, $\beta_{0e}/\beta_{0i}=\ex{2}$, $v_{A0i}/c=0.05$ and $m_i/m_e=16$. We find that only for $\omega_{0ci}/q\lesssim 200$ we can capture the physics of both ions and electrons with sufficient accuracy, and in \fig{mag} we show the cases $\omega_{0ci}/q=50$ (blue lines), $\omega_{0ci}/q=100$ (green) and $\omega_{0ci}/q=200$ (red). Yet, if we are only interested in the ion physics, we can extend our investigation to much larger values of $\omega_{0ci}/q$. This is shown in the subpanels of \fig{mag}(a), (d) and (e), where we study the development of the ion cyclotron instability up to the case $\omega_{0ci}/q=3200$, comparable to the range $\omega_{0ci}/q=1000-4000$ explored in the hybrid simulations by \citet{hellinger_05}.

As shown in \fig{mag}, the development of the ion cyclotron instability is similar, for different values of $\omega_{0ci}/q$. In all cases, the magnetic and electric energies grow exponentially (panels (d) and (e), respectively), with a growth rate that scales with the compression rate, rather than with the ion cyclotron frequency \citep[see also][for similar conclusions in a system dominated by the mirror instability]{riquelme_14}. The wavelength of the dominant mode is nearly insensitive to $\omega_{0ci}/q$, as shown in \fig{mag}(f), with only a marginal tendency for longer wavelengths at higher $\omega_{0ci}/q$. On the other hand, the growth of the unstable modes happens earlier for higher values of $\omega_{0ci}/q$. This trend is also apparent in \fig{mag}(a), where we show the ion anisotropy with solid lines. For faster compressions (i.e., smaller $\omega_{0ci}/q$, blue line), the system goes farther into the unstable region, before pitch-angle scattering off the growing ion cyclotron waves brings the ion anisotropy back to the threshold of marginal stability (shown with a black dashed line in \fig{mag}(a)). Since the anisotropy overshoot into the unstable region is more pronounced for smaller $\omega_{0ci}/q$, stronger magnetic fluctuations are needed to drive the system back to marginal stability, at smaller $\omega_{0ci}/q$ (see the magnetic energy at the end of the exponential phase in \fig{mag}(d)). In contrast, for slow compressions (red line in \fig{mag}(a)), the ion anisotropy does not move far from the threshold condition \citep[see also][for similar conclusions in a system dominated by the mirror instability]{riquelme_14}, and the amplitude of the ion cyclotron waves at the end of the exponential phase tends to be smaller (compare red and blue lines in \fig{mag}(d)).

In the subpanels of \fig{mag}(a), (d) and (e) we extend our analysis to larger values of $\omega_{0ci}/q$, towards the astrophysically-relevant limit $\omega_{0ci}/q\gg1$.\footnote{If electrons are non-relativistic and the mass ratio $m_i/m_e$ is much larger than unity, the electron physics will typically be in the asymptotic regime $\omega_{0ce}/q\gg1$ even for the moderate values of $\omega_{0ci}/q=50-200$ chosen in the main panels of \fig{mag}.} In these three subpanels, we plot on the horizontal axis the value of $\omega_{0ci}/q$ in the range from $50$ to $3200$, with each tick mark corresponding to an increment by a factor of two (so, logarithmic scale). In the subpanel of \fig{mag}(a), we show both the peak value of the ion anisotropy (filled black circles) and the time at which the ion anisotropy peaks (filled orange circles), which is a good proxy for the onset time of the ion cyclotron instability. In the subpanels of \fig{mag}(d) and (e), we show respectively the magnetic and electric energies in ion cyclotron waves normalized to the mean field energy, measured at the end of the exponential phase of the ion cyclotron instability. 

From the subpanel in \fig{mag}(a), we argue that both the onset time of the instability and the maximum ion anisotropy approach a constant value in the limit $\omega_{0ci}/q\gg1$, with no residual dependence on this parameter. More precisely, the onset time of the ion cyclotron waves, which is $\simeq 0.4\,q^{-1}$ for $\omega_{0ci}/q=100$, tends toward $0.25\,q^{-1}$ for $\omega_{0ci}/q\gg1$. Similarly, for the parameters adopted in \fig{mag}, the peak ion anisotropy approaches an asymptotic value of $\simeq 0.5$, a factor of two smaller than in our reference case $\omega_{0ci}/q=100$. On the other hand, \fig{mag}(a) demonstrates that at late times the ion anisotropy approaches the same threshold of marginal stability in \eq{margi}, regardless of $\omega_{0ci}/q$. For this reason, we argue that, in the absence of the electron whistler phase, the electron anisotropy during the growth of the ion cyclotron instability can still be scaled to the ion anisotropy at marginal stability $A_{i,\rm MS}$, as assumed in the last term of \eq{cond3}, but the coefficient of proportionality in the limit $\omega_{0ci}/q\gg1$ should be a factor of two smaller than in \eq{cond3}. The trend for a lower degree of electron anisotropy as $\omega_{0ci}/q$ increases is confirmed by the dotted lines of \fig{mag}(a) at $q\,t\gtrsim 0.6$.

With increasing $\omega_{0ci}/q$, the magnetic and electric energies in ion cyclotron waves also approach a constant value, as shown in the subpanels of \fig{mag}(d) and (e), respectively. The value of $\la \delta B_\perp^2\ra/|\la \bvec \ra|^2$ at the end of the exponential phase of ion cyclotron growth asymptotes to $\simeq 0.1$, roughly a factor of three smaller than in our standard case $\omega_{0ci}/q=100$ (subpanel in \fig{mag}(d)). The asymptotic value of $\la \delta E_\perp^2\ra/|\la \bvec \ra|^2$ is smaller than for $\omega_{0ci}/q=100$ by a similar factor (subpanel in \fig{mag}(e)). In summary, we conclude that the coefficient in \eq{cond1} should be reduced by a factor of three, for $\omega_{0ci}/q\gg1$.



In our analytical model of electron heating, based on Eqs.~\eqn{cond1}-\eqn{cond3} and Eqs.~\eqn{scal1}-\eqn{scal3}, we have not taken into account the explicit dependence on  $\omega_{0ci}/q$, which would change the coefficients in \eq{cond1} and \eq{cond3} in the way we have just described. This explains why our model of electron heating, which is benchmarked at $\omega_{0ci}/q=100$ (green line in \fig{mag}(c)), tends to overpredict the heating efficiency at higher values of $\omega_{0ci}/q$ (red line for $\omega_{0ci}/q=200$), whereas it underpredicts the results of our simulations at lower $\omega_{0ci}/q$ (blue line for $\omega_{0ci}/q=50$).

\begin{table}
\centering
\caption{Fractional Contributions to Electron Heating}\label{tab:mag}
\begin{tabular}{cccc}\hline\hline
Run & $\frac{\la\Delta\gamma_{\partial B}-|\Delta\gamma_{\rm curv}|\ra}{\la \Delta\gamma_{e,q}\ra}$ 
 & $\frac{\la\Delta\gamma_{\grad B}\ra}{\la \Delta\gamma_{e,q}\ra}$ 
 & $\frac{\la\Delta\gamma_{E\cross B}\ra}{\la \Delta\gamma_{e,q}\ra}$\\[4pt]
\hline\hline
$\omega_{0ci}/q=50$ & 0.49 & 0.09 & 0.42 \\\hline
$\omega_{0ci}/q=100$ &0.49 & 0.09 &  0.42 \\\hline
$\omega_{0ci}/q=200$ &0.49 & 0.09 & 0.42 \\\hline\hline
\multicolumn{4}{l}{%
  \begin{minipage}{8cm}%
    Note: We fix $\beta_{0i}=20$, $\beta_{0e}/\beta_{0i}=\ex{2}$, $v_{A0i}/c=0.05$ and $m_i/m_e=16$.%
  \end{minipage}%
}\\
\end{tabular}
\end{table}
\vspace{0.35in}

\begin{figure*}[tbp]
\begin{center}
\includegraphics[width=1\textwidth]{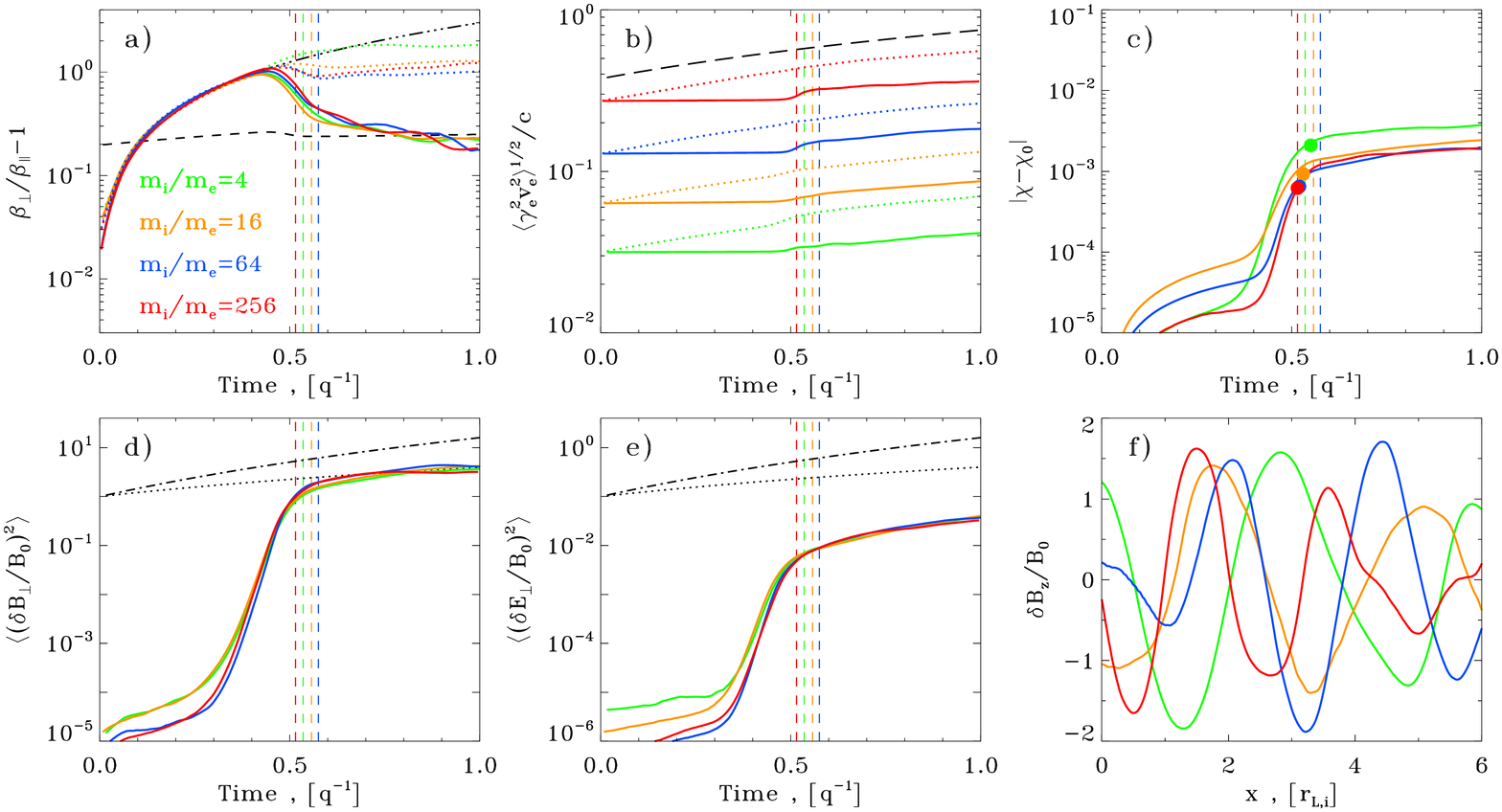}
\caption{Temporal and spatial development of compression-driven instabilities, for $T_{0e}/T_{0i}=\ex{2}$ and different choices of the mass ratio $m_i/m_e$, as detailed in the legend of panel (a). We fix  $\beta_{0i}=20$, $v_{A0i}/c=0.05$ and $\omega_{0ci}/q=100$. See the caption of \fig{tratio} for details. }
\label{fig:mimelow}
\end{center}
\end{figure*}

\begin{figure*}[tbp]
\begin{center}
\includegraphics[width=1\textwidth]{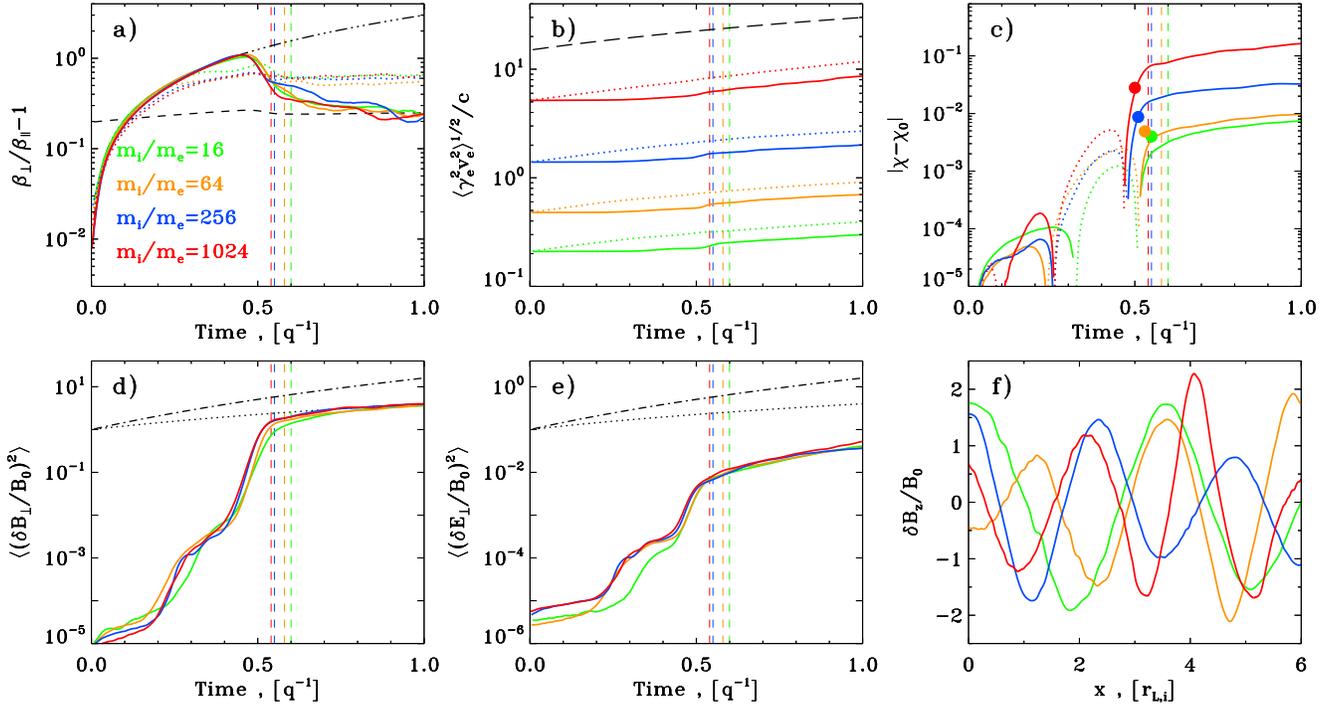}
\caption{Temporal and spatial development of compression-driven instabilities, for $T_{0e}/T_{0i}=\ex{1}$ and different choices of the mass ratio $m_i/m_e$, as detailed in the legend of panel (a). We fix  $\beta_{0i}=20$, $v_{A0i}/c=0.05$ and $\omega_{0ci}/q=100$. See the caption of \fig{tratio} for details.}
\label{fig:mimehigh}
\end{center}
\end{figure*}

\vspace{-0.15in}
\subsection{Dependence on the Mass Ratio}\label{sec:mime}
In this section, we investigate the dependence of our results on the mass ratio $m_i/m_e$, for two representative cases of the electron-to-ion temperature ratio, $T_{0e}/T_{0i}=\ex{2}$ in \fig{mimelow} and $T_{0e}/T_{0i}=\ex{1}$ in \fig{mimehigh}. In both cases, we fix $\beta_{0i}=20$, $v_{A0i}/c=0.05$ and $\omega_{0ci}/q=100$.

For both $T_{0e}/T_{0i}=\ex{2}$ in \fig{mimelow} and $T_{0e}/T_{0i}=\ex{1}$ in \fig{mimehigh}, the development of the ion cyclotron instability does not depend on the mass ratio \citep[see also][for similar conclusions in a system dominated by the mirror instability]{riquelme_14}. The curves that describe the evolution of the magnetic and electric energy densities in ion cyclotron waves  nearly overlap at $q\, t\gtrsim 0.4$ (panels (d) and (e), respectively), which suggests that the electron physics does not directly affect the growth of the ion cyclotron instability. Similarly, the dominant wavelength of ion cyclotron modes, normalized to the ion Larmor radius as in panel (f), does not depend on the mass ratio. On the other hand, the choice of mass ratio has an influence on the electron anisotropy (dotted lines in panel (a)) and on the efficiency of electron heating (panel (c)), as we now explain in detail.

For $T_{0e}/T_{0i}=\ex{2}$ in \fig{mimelow}, we find that the electron heating efficiency --- as quantified by the parameter $\chi$ in \fig{mimelow}(c) --- drops as the mass ratio increases from $m_i/m_e=4$ (green line) to $m_i/m_e=16$ (orange line), and it remains constant for higher mass ratios (blue for $m_i/m_e=64$ and red for $m_i/m_e=256$). This trend corresponds to a change in the dominant mechanism of electron heating. For mass ratios as small as $m_i/m_e=4$, most of the electron heating comes from the E-cross-B term in \eq{scal3} (see the first row in \tab{mimelow}). This term depends on the mass ratio as $\propto m_e/m_i$ (see \eq{scal3}), so it gives poorer heating efficiencies for higher mass ratios, which explains the trend in \fig{mimelow}(c). On the other hand, if $m_i/m_e=16$ or higher, the physics of electron heating is primarily controlled by the combination of \eq{scal1} and \eq{scal2} (see \tab{mimelow}), which are both independent of $m_i/m_e$. In turn, this explains why the heating efficiency is independent of mass ratio, for $m_i/m_e\ge16$.\footnote{We remark that this conclusion only applies to non-relativistic electrons. If the electrons are ultra-relativistic, at fixed $T_{0e}/T_{0i}$ the mean electron Lorentz factor $\la \gamma_{0e}\ra$ that appears in the change of the $\chi$ parameter (\eq{delchieff}) will scale as $\la \gamma_{0e}\ra\propto m_i/m_e$.}

In all cases, the results of our simulations (solid lines in \fig{mimelow}(c)) are in excellent agreement with the predictions of our analytic model (filled circles in \fig{mimelow}(c)). The electron anisotropy in \fig{mimelow}(a) (dotted lines) provides support for one of the fundamental assumptions in our model, \eq{cond3}. In fact, we confirm that in all the cases where electron heating is controlled by terms that depend on the electron thermal content (i.e., Eqs.~\eqn{scal1} and \eqn{scal2}), the electron anisotropy during the growth of the ion cyclotron instability stays at $\sim 5\,A_{i,\rm MS}$ independently of the mass ratio. It follows that \eq{cond3} can be confidently employed in \eq{heat1} to obtain \eq{scal1}. In \fig{mimelow}(a), the only case that departs from this prediction is $m_i/m_e=4$ (dotted green line), where indeed the electron heating process is different, being dominated by the E-cross-B term (see the first row in \tab{mimelow}).

For a higher electron-to-ion temperature ratio ($T_{0e}/T_{0i}=\ex{1}$ in \fig{mimehigh}), the growth of the ion cyclotron mode at $q\,t\simeq 0.5$ follows a phase dominated by electron whistler waves (see the minor bump at $q\,t\simeq0.3$ in the magnetic and electric energy of panels (d) and (e), respectively). In this case, the electron anisotropy (dotted lines in \fig{mimehigh}(a)) stays at the threshold of marginal stability $A_{e,\rm MS}$ of the electron whistler mode, see \eq{marge}. This justifies our {\it ansatz} in \eq{cond3}, in the case $A_{e,\rm MS}\leq 5 \,A_{i,\rm MS}$. Pitch-angle scattering off whistler waves constrains the electron anisotropy both at $0.3\lesssim q\,t\lesssim 0.5$, during the whistler phase, and also at $q\,t\gtrsim 0.5$, after the growth of the ion cyclotron instability. In fact, the minor wiggles superimposed over the ion cyclotron waves in \fig{mimehigh}(f) (e.g., see the orange line at $x/r_{L,i}\lesssim2$) demonstrate that the whistler instability continues to operate at late times, on top of the dominant ion cyclotron mode.

As regard to electron heating for $T_{0e}/T_{0i}=\ex{1}$ (see \fig{mimehigh}(c)), we find that the heating efficiency, as quantified by the $\chi$ parameter, is a monotonic function of $m_i/m_e$. Here, the physics of electron heating is dominated by the terms in Eqs.~\eqn{scal1} and \eqn{scal2}. In fact, \tab{mimehigh} shows that the contribution of the E-cross-B term is always less than $10\%$, for the parameters explored in \fig{mimehigh}. Eqs.~\eqn{scal1} and \eqn{scal2} do not explicitly depend on the mass ratio. Yet, at fixed $T_{0e}/T_{0i}$, the mean electron Lorentz factor $\la \gamma_{0e}\ra$ that appears in the increase of the $\chi$ parameter (\eq{delchieff})
will scale with mass ratio as $\la \gamma_{0e}\ra\propto m_i/m_e$, as soon as the electrons become ultra-relativistic. This explains why the increase in the $\chi$ parameter  for the cases $m_i/m_e=16$ and $64$ is nearly identical (green and orange solid lines in \fig{mimehigh}(c), respectively), since the electrons are still non-relativistic. For $m_i/m_e=256$ and $1024$, the electrons are ultra-relativistic, and the change in the $\chi$ parameter scales as $\Delta \chi\propto\la \gamma_{0e}\ra\propto m_i/m_e$.

Finally, we point out that our model tends to overestimate the actual increase in the $\chi$ parameter for $m_i/m_e=16$ and 64 (compare the green and orange lines with the corresponding filled circles in \fig{mimehigh}(c)). As we discuss in \app{cool}, this is due to the energy lost by electrons to drive the growth of electron whistler waves.  The value of $\chi-\chi_0$ becomes negative (dotted lines in \fig{mimehigh}(c)), so that the subsequent increase driven by the ion cyclotron instability falls short of our analytical prediction (which does not take into account the degree of electron cooling during the whistler phase). A similar effect has been discussed in \sect{tratio}.

\begin{table}
\centering
\caption{Fractional Contributions to Electron Heating}\label{tab:mimelow}
\begin{tabular}{cccc}\hline\hline
Run & $\frac{\la\Delta\gamma_{\partial B}-|\Delta\gamma_{\rm curv}|\ra}{\la \Delta\gamma_{e,q}\ra}$ 
 & $\frac{\la\Delta\gamma_{\grad B}\ra}{\la \Delta\gamma_{e,q}\ra}$ 
 & $\frac{\la\Delta\gamma_{E\cross B}\ra}{\la \Delta\gamma_{e,q}\ra}$\\[4pt]
\hline\hline
$m_i/m_e=4$ & 0.22 & 0.04 & 0.74 \\\hline
$m_i/m_e=16$ &0.49 & 0.09 &  0.42 \\\hline
$m_i/m_e=64$ &0.72 & 0.13 & 0.15 \\\hline
$m_i/m_e=256$ &0.81 & 0.14 & 0.05 \\\hline\hline
\multicolumn{4}{l}{%
  \begin{minipage}{8cm}%
    Note: We fix $\beta_{0i}=20$, $\beta_{0e}/\beta_{0i}=\ex{2}$, $v_{A0i}/c=0.05$ and $\omega_{0ci}/q=100$.%
  \end{minipage}%
}\\
\end{tabular}
\end{table}
\vspace{0.35in}

\begin{table}
\centering
\caption{Fractional Contributions to Electron Heating}\label{tab:mimehigh}
\begin{tabular}{cccc}\hline\hline
Run & $\frac{\la\Delta\gamma_{\partial B}-|\Delta\gamma_{\rm curv}|\ra}{\la \Delta\gamma_{e,q}\ra}$ 
 & $\frac{\la\Delta\gamma_{\grad B}\ra}{\la \Delta\gamma_{e,q}\ra}$ 
 & $\frac{\la\Delta\gamma_{E\cross B}\ra}{\la \Delta\gamma_{e,q}\ra}$\\[4pt]
\hline\hline
$m_i/m_e=16$ &0.71 & 0.19 &  0.10 \\\hline
$m_i/m_e=64$ &0.76 & 0.21 & 0.03 \\\hline
$m_i/m_e=256$ &0.78 & 0.22 & $<0.01$ \\\hline
$m_i/m_e=1024$ &0.78 & 0.22 & $<0.01$ \\\hline\hline
\multicolumn{4}{l}{%
  \begin{minipage}{8cm}%
    Note: We fix $\beta_{0i}=20$, $\beta_{0e}/\beta_{0i}=\ex{1}$, $v_{A0i}/c=0.05$ and $\omega_{0ci}/q=100$.%
  \end{minipage}%
}\\
\end{tabular}
\end{table}
\vspace{0.35in}


\section{Summary and Discussion}\label{sec:summary}
In this work, the second of a series, we have investigated by means of PIC simulations how the efficiency of electron heating by ion velocity-space instabilities (more specifically, the ion cyclotron instability) depends on the physical conditions in low-luminosity two-temperature accretion flows. Pressure anisotropies are continuously generated in collisionless accretion flows due to the fluctuating magnetic fields associated with the non-linear stages of the magnetorotational instability (MRI, \citealt{balbus91,balbus98}). Field amplifications induced by the MRI, coupled to the adiabatic invariance of the magnetic moments of charged particles, drive a temperature anisotropy $P_\perp>P_\parallel$ (relative to the local field), which relaxes via velocity-space instabilities. 

In Paper I, we have developed a fully-kinetic method for studying velocity-space instabilities in a system where the field is continuously amplified. So, the anisotropy is constantly driven (as a result of the field amplification), rather than assumed as a prescribed initial condition, as in most earlier works. In our setup, the increase in magnetic field is driven by {\it compression}, but our results hold regardless of what drives the field amplification, so they can be equally applied to the case where velocity-space instabilities are induced by incompressible {\it shear} motions \citep[as in][]{riquelme_14}.

In Paper I we found that, for the values of ion plasma beta $\beta_{0i}\sim 5-30$ expected in the midplane of low-luminosity accretion flows \citep[e.g.,][]{sadowski_13}, the dominant mode for $T_{0e}/T_{0i}\lesssim 0.2$ is the ion cyclotron instability, rather than the mirror instability. Since the wavevector of the ion cyclotron instability is aligned with the mean magnetic field, the relevant physics can be conveniently studied by means of 1D simulations, with the box aligned with the ordered field. 
In this work, we have studied with 1D PIC simulations the efficiency of electron heating  by the ion cyclotron instability. Fully-kinetic simulations with mobile ions and a realistic
mass ratio are required to capture at the same time electron-scale and ion-scale instabilities (in our setup, the electron whistler and ion cyclotron modes), as well as to properly describe the energy transfer from ions to electrons. 

We have assessed the dependence of the electron heating
efficiency on the initial ratio
between electron and proton temperatures $T_{0e}/T_{0i}$, on the ion beta $\beta_{0i}$ (namely, the ratio of ion thermal pressure to magnetic pressure), on the Alfv\'en speed $v_{Ai0}$, on the compression rate $q$ (in units of the proton cyclotron frequency $\omega_{0ci}$), and on the proton to
electron mass ratio $m_i/m_e$, which we have explored up to realistic values. Our analysis does consistently allow for relativistic temperatures and thus it is possible to study the limit of non-relativistic ions and ultra-relativistic electrons that is of particular interest for two-temperature disk models.

Eqs.~\eqn{scal1}-\eqn{scal3} emphasize how the various contributions to electron heating --- whose physical origin has been described in Paper I --- depend on flow conditions. Their sum gives the overall electron energy gain associated with the development of the ion cyclotron instability, and it represents the main result of this paper. Eqs.~\eqn{scal1}-\eqn{scal3} show the explicit dependence of the electron energy gain on $T_{0e}/T_{0i}$, $\beta_{0e}$, $\beta_{0i}$ and $m_i/m_e$, as based on the analytical model presented in Paper I and on the scalings in Eqs.~\eqn{cond1}-\eqn{cond3}, that we have justified analytically (see \app{sat}) and extensively validated with PIC simulations in this work. In addition, we find that our results do not explicitly depend on the Alfv\'en velocity $v_{A0i}$, as long as it is non-relativistic, and are weakly dependent on $\omega_{0ci}/q$, as long as $\omega_{0ci}/q\gg1$ ($\omega_{0ci}/q\sim 10^7$ in accretion flows, assuming that the timescale of turbulent eddies is comparable to the orbital time). The same weak dependence on  $\omega_{0ci}/q\gg1$  has been found in the shear-driven PIC simulations of velocity-space instabilities in electron-positron plasmas of \citet{riquelme_14}, as regard to the mirror instability.

Another important result of our work concerns the ion response to the ion cyclotron instability. We have assessed that, in our case of driven ion cyclotron instability, the magnetic energy in ion cyclotron waves in the saturated stage scales with the ion beta as $\la \delta B_{\perp}^2\ra/|\bavg|^2\sim0.07\, \beta_{0i}^{0.5}$. Pitch-angle scattering off the ion cyclotron waves maintains the ion anisotropy at the threshold of marginal stability $[\beta_\perpi/\beta_\pari-1]_{\rm MS}\simeq\coeffi/\beta_\pari^\powi$. Similarly, in the cases when the electron anisotropy exceeds the threshold for the electron whistler instability before the onset of the ion cyclotron mode, the electron anisotropy is constrained to follow the track of marginal stability $[\beta_\perpe/\beta_\pare-1]_{\rm MS}\simeq\coeffe/\beta_\pare^\powe$ by efficient pitch-angle scattering off the electron whistler waves. These scalings had been widely investigated in the context of undriven velocity-space instabilities (i.e., where the anisotropy is prescribed as an initial condition, see \citet{gary_book} for  a review), yet never with fully-kinetic simulations having mobile ions and a realistic
mass ratio, as we employ in this work. In this work, we have estimated such scalings in the case that the ion cyclotron instability is induced by a continuous field amplification.

Our work has implications for the electron physics in two-temperature accretion flows. The electron energy gain associated with the growth of the ion cyclotron instability (see Eqs.~\eqn{scal1}-\eqn{scal3}) can be incorporated in GRMHD simulations of low-luminosity accretion flows, by adding a source term to the electron thermal evolution.
When coupled to cooling by radiation, this will provide a physically-grounded model for assessing the two-temperature nature of low-luminosity accretion flows like Sgr A$^*$ \citep[e.g.,][]{yuan_narayan_14}. In this sense, our results provide solid evidence that the ion cyclotron instability has a tendency to
 equilibrate the ion and electron temperatures, in the regime $T_{0e}/T_{0i}\lesssim 0.2$ where the ion cyclotron instability dominates over the mirror mode.
 
The physics of electron heating at higher electron-to-ion temperature ratios (i.e., $T_{0e}/T_{0i}\gtrsim 0.2$) is  beyond the scope of this work. In this regime, 2D simulations would be needed to assess the efficiency of electron heating by the mirror instability, whose wavevector is oblique with respect to the mean field. On the other hand, at such high electron temperatures ($T_{0e}/T_{0i}\gtrsim 0.3$), the threshold for the electron whistler instability is exceeded before the onset of the mirror mode, for the range of $\beta_{0i}$ expected in accretion flows. As we have found in this work, the growing electron whistler waves will remove thermal energy from the electron population, resulting in net cooling. The relative role of electron cooling by the whistler instability and electron heating by the mirror mode at $T_{0e}/T_{0i}\gtrsim 0.2$ should be investigated with 2D simulations having $\omega_{0ci}/q\gg1$, and it is deferred to a future work. It will be needed to complement the findings of this paper, which apply to $T_{0e}/T_{0i}\lesssim 0.2$.

Another important question that we have not addressed in this work is the generation of non-thermal electrons, whose emission is required for modeling the broad-band signature of Sgr A$^*$ \citep{yuan+03,yuan_narayan_14,lynn_14}. Particle acceleration due to the development of anisotropy-driven instabilities is generally believed to be quite inefficient \citep[e.g.,][]{kennel_66}, whereas other mechanisms --- most notably, magnetic reconnection --- might be more promising. PIC simulations of carefully designed systems will be needed to investigate the origin of non-thermal electrons in low-luminosity accretion flows.

\acknowledgements
We thank R. Narayan for useful discussions. L.S. is supported by NASA through Einstein
Postdoctoral Fellowship grant number PF1-120090 awarded by the Chandra
X-ray Center, which is operated by the Smithsonian Astrophysical
Observatory for NASA under contract NAS8-03060. This work is supported in part by NASA via the TCAN award grant number NNX14AB47G. L.S. gratefully acknowledges access to the PICSciE-OIT High Performance Computing Center and Visualization
Laboratory at Princeton University. The simulations were also performed on XSEDE resources under
contract No. TG-AST120010, and on NASA High-End Computing (HEC) resources through the NASA Advanced Supercomputing (NAS) Division at Ames Research Center. 

\appendix
\section{A. The Electromagnetic Fields at Saturation}\label{sec:sat}
In this appendix, we assess how the saturation amplitude of the electromagnetic fields (i.e., at the end of the phase of exponential growth) depends on flow conditions, for both the ion cyclotron instability and the electron whistler instability. As we show below, the physics of field saturation is different in our setup, where the anisotropy is constantly driven, with respect to the case of undriven systems explored by, e.g., \citet{gary_93,gary_93d}. Our goal is only to provide approximate scalings, and we refer to, e.g., \citet{gary_feldman_78} or \citet{gary_85} (see also \citealt{hamasaki_73,yoon_92,hellinger_09,hellinger_13}) for a detailed analysis of the evolution of the electromagnetic fields beyond the exponential phase.

At the end of the exponential phase, the magnitude of the magnetic fields has to be large enough such that efficient pitch-angle scattering drives the particle anisotropy back to marginal stability. In other words, the scattering frequency $\nu_{\rm scatt}$ in the fields of the unstable modes will be counter-acting the overall effect of field amplification, which feeds the anisotropy. Following \citet{braginskii_65} (see also \citealt{scheko_08,rosin_11,kunz_14}), it is required that
\be\label{eq:appb1}
\nu_{\rm scatt}\sim \frac{1}{\beta_{\perp}/\beta_{\parallel}-1} \frac{\ud \ln|\bavg|}{\ud t}~~~,
\ee
where $|\bavg|\propto \qt^2$ is the magnitude of the ordered field. The scattering is provided by pitch-angle diffusion in the electromagnetic fields of the growing instability (e.g., \citealt{kennel_66,blandford_eichler_87,devine_95}), so that
\be\label{eq:appb2}
\nu_{\rm scatt}\sim \left(\frac{\lambda}{r_{L,\delta B}}\right)^2 \frac{v}{\lambda}~~~,
\ee
where $\lambda$ is the wavelength of the relevant instability, $r_{L,\delta B}$ is the characteristic particle Larmor radius in the  magnetic field $\la\delta B_{\perp}^2\ra^{1/2}$ of the unstable waves and $v$ is the characteristic particle velocity. In the expression above, $ (r_{L,\delta B}/\lambda)^2\gg1$  is the number of scatterings required for a deflection of $\sim \pi/2$, whereas $\lambda/v$ is the characteristic duration of each scattering event.

\eq{appb1} and \eq{appb2} hold for both the ion cyclotron instability and the electron whistler instability. We first focus on the ion cyclotron instability. At marginal stability, its characteristic wavelength and  phase speed are respectively $\lambda_i\simeq2\pi A_{i,\rm MS}^{-1} \,c/\omega_{\rm pi}$ and $\omega_i/k_i\simeq v_{Ai}$ \citep[e.g.,][]{kennel_66,davidson_75,yoon_92,yoon_10}, where we have taken the limit $A_{i,\rm MS}\simeq\coeffi/\beta_\pari^\powi\ll1$ appropriate for $\beta_i\gg1$. For the parameters relevant in accretion flows, ions will be non-relativistic. Their characteristic velocity is the thermal speed $\la v_{i}\ra\simeq\sqrt{3 \,k_B\, T_i/m_i}$, and their Larmor radius in the turbulent fields is $r_{Li,\delta B}\simeq m_i \la v_{i}\ra c/e\la\delta B_{\perp}^2\ra^{1/2}$.\footnote{Since the characteristic wavelength of the ion cyclotron mode is $\lambda_i\sim A_{i, \rm MS}^{-1} \,c/\omega_{\rm pi}\sim \beta_{i}^{\powi}\,c/\omega_{\rm pi}$ and  the Larmor radius in the turbulent fields is $r_{Li,\delta B}\sim\beta_i^{1/2}\,|\bavg|\, \la\delta B_{\perp}^2\ra^{-1/2}\,c/\omega_{\rm pi}$, it is straightforward to derive that the assumption  $ (r_{Li,\delta B}/\lambda_i)^2\gg1$ implicit in \eq{appb2} is equivalent to $\la\delta B_{\perp}^2\ra/|\bavg|^2\ll1$, which is indeed satisfied for the parameters explored in this work.}
By equating \eq{appb1} and \eq{appb2}, we find
\be\label{eq:appb3}
\frac{\la\delta B_{\perp}^2\ra}{|\bavg|^2}\propto \beta_{0i}^{1/2}~~~.
\ee
Here, we have neglected any explicit time dependence (e.g., we have used the initial ion beta $\beta_{0i}$), since our goal is to assess the approximate scalings of the fields at the end of the exponential phase, and not to describe the subsequent secular phase.
From Maxwell's equations, the magnitude of the electric fields at saturation will be
\be\label{eq:appb4}
\frac{\la\delta E_{\perp}^2\ra}{|\bavg|^2}\sim \left(\frac{\omega_i}{k_i c}\right)^2\frac{\la\delta B_{\perp}^2\ra}{|\bavg|^2}\propto \frac{v_{A0i}^2}{c^2}\beta_{0i}^{1/2}~~~.
\ee
where $v_{A0i}$ is the Alfv\'en velocity at the initial time. The scalings in \eq{appb2} and \eqn{appb3} have been extensively checked in the main body of the paper. 

A similar argument can be put forward for the electron whistler instability. At marginal stability, the characteristic wavelength and phase speed are respectively $\lambda_e\simeq 2\pi A_{e,\rm MS}^{-1/2}\,c/\omega_{\rm pe}$ and $\omega_e/k_e\simeq c\,A_{e,\rm MS}^{1/2}\,(A_{e,\rm MS}+1)^{-1}\, \omega_{ce}/\omega_{\rm pe}$ \citep{kennel_66,ossakow_72,ossakow_72b,yoon_87,yoon_11,bashir_13}, where the electron anisotropy at marginal stability is $A_{e,\rm MS}\simeq\coeffe/\beta_\pare^\powe$. Since we account for the possibility that electrons are ultra-relativistic, the proper definitions of the electron cyclotron frequency and plasma frequency are $\omega_{ce}=e|\bavg|/\la\gamma_e\ra m_e c$ and $\omega_{\rm pe}=\sqrt{4 \pi n e^2/\la \gamma_e \ra m_e}$, respectively smaller by a factor of $\la \gamma_e\ra$ and of $\la \gamma_e\ra^{1/2}$ than the non-relativistic formulae. The electron Larmor radius in the turbulent fields will be $r_{Le,\delta B}\simeq \la p_{e}\ra c/e\la\delta B_{\perp}^2\ra^{1/2}$, where $\la p_{e}\ra$ is the characteristic electron thermal momentum. By equating \eq{appb1} and \eq{appb2}, we obtain
\be\label{eq:appb5}
\frac{\la\delta B_{\perp}^2\ra}{|\bavg|^2}\propto \sqrt{\frac{\beta_{0e}}{A_{e,\rm MS}}}\frac{m_e\la \gamma_{0e}\ra}{m_i}~~~,
\ee
where the electron anisotropy should be evaluated at marginal stability, so $A_{e,\rm MS}\propto \beta_{0e}^{-\powe}$. We notice that the strength of the fields resulting from the electron whistler instability is independent of mass ratio (everything else being fixed) only if the initial electron temperature is $k_BT_{0e}/m_e c^2\gg1$ (i.e., ultra-relativistic electrons), so that the electron Lorentz factor is $\la \gamma_{0e}\ra\simeq 3\,k_BT_{0e}/m_e c^2$. 
The energy in the electric fields will be
\be
\frac{\la\delta E_{\perp}^2\ra}{|\bavg|^2}\,&\,\sim& \left(\frac{\omega_e}{k_e c}\right)^{2}\frac{\la\delta B_{\perp}^2\ra}{|\bavg|^2}\\&\propto &\frac{v_{A0e}^2}{c^2}\frac{\sqrt{A_{e,\rm MS}\,\beta_{0e}}}{(A_{e,\rm MS}+1)^2}\frac{m_e\la \gamma_{0e}\ra}{m_i}~~,\label{eq:appb6}
\ee
where the electron Alfv\'en velocity at the initial time is defined as $v_{A0e}/c\equiv \omega_{0ce}/\omega_{0\rm pe}=|\bmath{B}_0|/\sqrt{4 \pi \la \gamma_{0e}\ra n_0 m_e c^2}$.\footnote{Clearly, our discussion only applies if $v_{A0e}/c\ll1$. Otherwise, the proper definition of the Alfv\'en speed will be $v_{A0e}/c=\sqrt{\sigma_{0e}/(1+\sigma_{0e})}$, where $\sigma_{0e}=|\bmath{B}_0|^2/4\pi (n_0 m_e c^2+w_{0e})$ is the electron effective magnetization, and $w_{0e}$ is the electron enthalpy per unit volume measured at the initial time.} We have explicitly checked the scalings in \eq{appb5} and \eq{appb6} by running a set of dedicated simulations of the compression-driven whistler instability, in which the ion physics was artificially suppressed by considering infinitely massive ions. Our results will be reported elsewhere.

\section{B. Electron Cooling by the Whistler Instability}\label{sec:cool}
During the development of the electron whistler instability, the free energy available in the compression-induced electron anisotropy mediates the growth of whistler waves. The electric component of the waves is sub-dominant with respect to the magnetic component, as shown in \fig{tratio} and \fig{mimehigh} (compare panels (d) and (e)). The growth of the wave magnetic energy will result in cooling of the electron population. The fraction of electron energy lost during the exponential growth of the whistler instability can be estimated from \eq{appb5}, assuming energy conservation. This is a good approximation during the exponential phase of the instability, which is much shorter than the characteristic compression time $\sim q^{-1}$, so one can neglect the energy injected into the system by compression.

However, the energy content of the electron population --- as quantified by the $\chi$ parameter in \fig{tratio}(c) --- keeps decreasing after the exponential growth has terminated. In this Appendix, we demonstrate that such apparent electron cooling, and the associated characteristic evolution of the $\chi$ parameter (see the dotted lines in \fig{tratio}(c) and \fig{mimehigh}(c)) is just a result of the requirement that the electron distribution stays at marginal stability, after the end of the exponential phase of whistler growth.

Let us consider the equation for the evolution of the mean electron energy\footnote{In this section, for ease of notation we neglect the subscript ``e'' to indicate electrons.}
\be\label{eq:engain}
\frac{\ud \la \gamma-1\ra c^2}{\ud t}= \frac{2\,q}{\qt}\,\la \gamma v_{\perp}^2\ra-\frac{e}{m_e}\la\evec\cdot\bmath{v}\ra~~~.
\ee
After the exponential phase of the whistler instability, the right hand side of \eq{engain} is dominated by the compression term. Since $(\gamma-1)c^2=\gamma^2v^2/(\gamma+1)$, 
\be\label{eq:appc2}
\left\langle\frac{\gamma}{\gamma+1}\frac{\ud\, \gamma v^2}{\ud t}\right\rangle=\frac{2\,q}{\qt}\,\la \gamma v_{\perp}^2\ra~~,
\ee
where we have neglected the term with $\ud[\gamma/(\gamma+1)]/\ud t$ since it vanishes for both non-relativistic and ultra-relativistic electrons. At the threshold of marginal stability for the electron whistler instability we have 
\be
\frac{\la \gamma v_{\perp}^2\ra}{\la \gamma v_{\parallel}^2\ra}-1\simeq\frac{\coeffe}{\beta_{e\parallel}^\powe}~~,
\ee
which, in the simplifying limit $\beta_{e\parallel}\gg1$ (which, however, is not always realized for our parameters), yields $\la  \gamma v_{\perp}^2\ra\simeq  \la  \gamma v_{\parallel}^2\ra\simeq  \la  \gamma v^2\ra/3$. From \eq{appc2}, we then find that $\la \gamma v^2\ra\propto \qt^{4/3}$ in the non-relativistic limit $\la\gamma-1\ra\ll1$. In the ultra-relativistic limit $\la\gamma\ra\gg1$, we obtain $\la \gamma v^2\ra\propto \qt^{2/3}$. The results in the two opposite limits can be condensed in a unique scaling for the average electron momentum, such that $\la \gamma^2 v^2\ra\propto \qt^{4/3}$ for both non-relativistic and ultra-relativistic electrons.\footnote{As a side note, we remark that the scalings we have just found should also describe the evolution of the ion population after the saturation of the ion cyclotron instability.}

So far, we have described the evolution of the electron population, after the saturation of the electron whistler instability, but before the onset of the ion cyclotron instability. Thus, the ions are still evolving according to compression alone, so that $\la p_i^2\ra=\la p_{0i}^2\ra [1+2\,\qt^2]/3$. On the other hand, the electrons, which are constrained to remain at marginal stability, will evolve as $\la p_e^2\ra=\la p_{\rm exp}^2\ra \qt^{4/3}/(1+q \,t_{\rm exp})^{4/3}$, where the subscript ``exp'' refers here to the end of the exponential phase of whistler growth. It follows that the $\chi$ parameter, after the saturation of the whistler instability and before the growth of the ion cyclotron mode, will evolve as
\be\label{eq:appc4}
\chi\propto\frac{\qt^{4/3}}{1+2\,\qt^2}~~,
\ee
which, in the limit $\qt\gtrsim1$, results in apparent cooling towards $\chi\rightarrow0$. If the whistler instability grows much earlier than the ion cyclotron instability, the parameter $\chi-\chi_0$ will then saturate at $-\chi_0$. We have explicitly checked this prediction, as well as the temporal scaling in \eq{appc4}, by running a set of dedicated simulations of the compression-driven whistler instability, where the ions are artificially chosen to have infinite mass, so that they cannot participate in the electron whistler instability. Our results will be reported elsewhere.

\bibliography{bimax}
\end{document}